\documentclass[]{aa}  

\usepackage{makecell}
\usepackage{graphicx}
\usepackage{xcolor}
\usepackage{ulem}
\usepackage{txfonts}
\usepackage[colorlinks=true,urlcolor=blue,citecolor=blue,linkcolor=blue,breaklinks=true]{hyperref}

\usepackage{etoolbox}


\begin{document}

   \title{New mid-infrared imaging constraints on companions and protoplanetary disks around six young stars\thanks{Based on observations collected at the European Southern Observatory under ESO programmes 0101.C-0580(A), 60.A-9107(G), and 60.A-9107(N).}}
   \titlerunning{New Mid-IR Constraints on PPDs and Companions of 6 Young Stars}

   \author{D. J. M. Petit dit de la Roche\inst{1}
    \and N. Oberg\inst{2}
    \and M. E. van den Ancker\inst{1}
    \and I. Kamp\inst{2}
    \and R. van Boekel\inst{3}
    \and D. Fedele\inst{4}
    \and V. D. Ivanov\inst{1}
    \and M. Kasper\inst{1}
    \and H. U. K\"{a}ufl\inst{1}
    \and M. Kissler-Patig\inst{5}
    \and P. A. Miles-P\'{a}ez\inst{1}
    \and E. Pantin\inst{6}
    \and S. P. Quanz \inst{7}
    \and Ch. Rab\inst{2,8}
    \and R. Siebenmorgen\inst{1}
    \and L. B. F. M. Waters\inst{9,10}
          }

   \institute{European Southern Observatory, Karl-Schwarzschild-Strasse 2, 85748 Garching, Germany \\\email{dominique.petit@eso.org} 
         \and 
             Kapteyn Astronomical Institute, University of Groningen, P.O. Box 800, 9700 AV Groningen, The Netherlands
         \and
            Max-Planck Institut f\"{u}r Astronomie, K\"{o}nigstuhl 17, 69117 Heidelberg, Germany
         \and
             INAF-Osservatorio Astrofisico di Arcetri, Largo E. Fermi 5, 50125 Firenze, Italy
         \and
            European Space Agency, Camino Bajo del Castillo, s/n., Urb. Villafranca del Castillo, 28692 Villanueva de la Ca\~{n}ada, Madrid, Spain
         \and 
            CEA, IRFU, DAp, AIM, Universit\'{e} Paris-Saclay, Universit\'{e} Paris Diderot, Sorbonne Paris Cit\'{e}, CNRS, F-91191 Gif-sur-Yvette, France
         \and
            Institute for Particle Physics and Astrophysics, ETH Zurich, Wolfgang-Pauli-Strasse 27, 8093 Zurich, Switzerland
         \and
            Max-Planck-Institut f\"{u}r extraterrestrische Physik, Giessenbachstrasse 1, 85748 Garching, Germany
         \and
            Department of Astrophysics/IMAPP, Radboud University Nijmegen, P.O.Box 9010, 6500 GL Nijmegen, The Netherlands
         \and
            SRON Netherlands Institute for Space Research, Sorbonnelaan 2, 3584 CA Utrecht, The Netherlands
             }

   \date{Received XXXX; accepted XXXX}

 
  \abstract
   {Mid-infrared (mid-IR) imaging traces the sub-micron and micron-sized dust grains in protoplanetary disks and it offers constraints on the geometrical properties of the disks and potential companions, particularly if those companions have circumplanetary disks.}
   {We use the VISIR instrument and its upgrade NEAR on the VLT to take new mid-IR images of five (pre-)transition disks and one circumstellar disk with proposed planets and obtain the deepest resolved mid-IR observations to date in order to put new constraints on the sizes of the emitting regions of the disks and the presence of possible companions.}
   {We derotated and stacked the data to find the disk properties. Where available, we compare the data to \textsc{ProDiMo} (\textbf{Pro}toplanetary \textbf{Di}sk \textbf{Mo}del) radiation thermo-chemical models to achieve a deeper understanding of the underlying physical processes within the disks. We applied the circularised point spread function subtraction method to find upper limits on the fluxes of possible companions and model companions with circumplanetary disks.} 
   {We resolved three of the six disks and calculated position angles, inclinations, and (upper limits to) sizes of emission regions in the disks, improving upper limits on two of the unresolved disks. In all cases the majority of the mid-IR emission comes from small inner disks or the hot inner rims of outer disks. We refined the existing \textsc{ProDiMo} HD 100546 model spectral energy distribution (SED) fit in the mid-IR by increasing the PAH abundance relative to the ISM, adopting coronene as the representative PAH, and increasing the outer cavity radius to 22.3 AU.  We produced flux estimates for putative planetary-mass companions and circumplanetary disks, ruling out the presence of planetary-mass companions with $L > 0.0028 \,L_{\odot}$ for $a > 180$\,AU in the HD 100546 system. Upper limits of 0.5\,mJy-30\,mJy are obtained at 8\,$\mu$m-12\,$\mu$m for potential companions in the different disks.  We rule out companions with $L > 10^{-2} \,L_{\odot}$ for $a > 60$\,AU in TW Hydra, $a > 110$\,AU in HD 169142, $a > 150$\,AU in HD 163296, and $a > 160$\,AU in HD 36112.}
   {The mid-IR emission comes from the central regions and traces the inner areas of the disks, including inner disks and inner rims of outer disks.
   Planets with mid-IR luminosities corresponding to a runaway accretion phase can be excluded from the HD 100546, HD 169142, TW Hydra, and HD 36112 systems at separations >1$\arcsec$. We calculated an upper limit to the occurrence rate of wide-orbit massive planets with circumplanetary disks of 6.2\% (68\% confidence). Future observations with METIS on the ELT will be able to achieve a factor of 10 better sensitivity with a factor of 5 better spatial resolution. MIRI on \textit{JWST} will be able to achieve 250 times better sensitivity. Both will possibly detect the known companions to all six targets.}


   \maketitle
%

\begin{table*}[ht]
    \centering
    \caption{\label{tab:targets} Stellar and disk properties of the target stars. Stellar masses, luminosities, and temperatures, where possible, have been taken from the DIANA models of the targets, which are fit to multiple data sets. \textit{(a)} \citet{Gaia2020}. \textit{(b)} \citet{Wichittanakom2020,Miley2019,Casassus2019, Jamialahmadi2018,Mendigutia2017,Pineda2014,Avenhaus2014,Walsh2014,Leinert2004}. \textit{(c)} \citet{Garufi2014, Wichittanakom2020}. \textit{(d)} \citet{Perez2019,Panic2008,Raman2006,vanBoekel2005}. \textit{(e)} \citet{Nayakshin2020, Sokal2018}. \textit{(f)} \citet{Wichittanakom2020,Rosotti2020,Yu2019,Long2017,Benisty2017,Wagner2015}. \textit{(g)} \citet{Isella2010, Meeus2012}.}
    \begin{tabular}{cccccccccc}
        \hline \hline
         Target & Age (Myr) & $M_*$ ($\mathrm{M_\odot}$) & $T$ (K) & $L_*$ ($\mathrm{L_\odot}$) & $d$ (pc)$^{\mathrm{(a)}}$ & $PA$ ($^\circ$) & $i$ ($^\circ$) & Structures & \makecell{Companions\\ detected} \\ \hline
         HD 100546$^{(b)}$ & 5 & 2.5 & 10600 & 30 & 108.1 & 140 & 46 & gap, spiral arms & 3 \\
         HD 163296$^{(c)}$ & 6 & 2.5 & 9000 & 34.7 & 101.0 & 137 & 43 & rings & 3 \\
         HD 169142$^{(d)}$ & 4-16 & 1.8 & 7800 & 9.8 & 114.9 & 5 & 13 & rings & 3\\
         TW Hydra$^{(e)}$ & 3-15 & 0.8 & 4000 & 0.2 & 60.1 & 150 & 7 & rings & 1 \\
         HD 100453$^{(f)}$ & 11 & 1.5 & - & 6 & 103.8 & 145 & 35 & gap, spiral arms & M dwarf\\
         HD 36112$^{(g)}$ & 4 & 2.0 & 8200 & 22 & 155.9 & 62 & 21 & \makecell{cavity, rings, \\clumps, spirals} & 2\\ \hline
    \end{tabular}
\end{table*}

\section{Introduction}

Transition disks are believed to represent an intermediate stage of planet formation between the protoplanetary disk and a gasless, fully formed planetary system. Scattered light imaging in the near-infrared (near-IR) and thermal sub-millimetre observations with ALMA have revealed detailed structures in many transition disks, including rings, spirals, and warps \citep[e.g.][]{Francis2020}. These features can be a result of the accretion of gas and dust onto a planet, although they can also be explained by other processes in the disk such as shadowing from the inner rim, snowlines, or hydrodynamic effects \citep[e.g.][]{Siebenmorgen2012,van_der_Marel_2018}. Studying transition disks is an important step in understanding planet formation.  
Mid-infrared (mid-IR) direct imaging traces dust of $\sim$150\,K in the disk. Additionally, the disk is expected to re-emit a large fraction of the stellar flux in the infrared \cite[e.g.][]{Dullemond2010}.  Mid-IR imaging can thus further constrain disk properties, especially when combined with observations at other wavelengths. It also allows us to search for thermal emission from (planetary) companions, especially if these companions still have circumplanetary disks (CPDs), which are expected to be bright in the mid-IR.\par

We used the VLT Imager and Spectrograph for the mid-InfraRed \citep[VISIR;][]{Lagage2004} and its upgraded version Near Earths in the AlphaCen Region \citep[NEAR;][]{Kasper2017} to obtain the deepest resolved mid-IR images of five Herbig Ae/Be (pre-)transition disks and one other circumstellar disk to date. The instruments that we used are more sensitive and the observation time is longer than in any previous studies \citep{Liu2003, vanBoekel2004, Leinert2004, Verhoeff2009, Panic2014, Marinas2011, Doucet2006, Honda2012, Okamoto2017, Maaskant2013, Ratzka2007, Arnold2012, Khalafinejad2016}. Additionally, the use of adaptive optics (AO) on NEAR provides us with better angular resolution and PSF stability. These new data allowed us to put new constraints on the disk and the presence of possible companions of each of the six targets. 

To contextualise the observations of our primary target HD 100546 and secondary targets, we used the radiation thermo-chemical disk modelling code \textsc{ProDiMo} (\textbf{Pro}toplanetary \textbf{Di}sk \textbf{Mo}del; see Sect.~\ref{section:prodimo}). Our HD 100546 disk model is the result of a multi-wavelength spectral energy distribution (SED) fit, which will allow us to compare the predicted and observed total flux within the observed bands \citep{Woitke2019}. Our synthetic images of the HD 100546 circumstellar disk enabled us to search for a non-axisymmetric disk structure. 
 The radiative transfer results allowed us to determine the mid-IR extinction along line-of-sights to the midplane and the resulting obscuration of putative embedded companions.  The disk modelling code can be applied further to produce SEDs for planetary companions and circumplanetary disks to compare theoretical fluxes with detection limits \citep{Rab_2019}.

Section \ref{sec:targets} describes the targets and in Sect.~\ref{sec:obs} we show the observations and the data analysis. The \textsc{ProDiMo} model is discussed in Sect.~\ref{section:prodimo} and compared to the data in Sect.~\ref{sec:comparison}. Limits on possible companions are discussed in Sect.~\ref{sec:companions} and for three of the targets planetary models with circumplanetary disks are analysed. Finally, our discussion and conclusions are presented in Sect.~\ref{sec:conclusions}.

\section{Targets}
\label{sec:targets}

The following targets were observed: HD 100546, HD 163296, HD 169142, TW Hydra, HD 100453, and HD 36112/MWC 758 (see Table~\ref{tab:targets}). 
These stars were selected to study the influence of features such as spiral arms, circular gaps, and inner cavities, seen in near-IR scattered light images on the mid-IR morphology of the disk which is dominated by thermal emission.

All six targets are young disks with ages of 3-16\,Myr and, with the exception of HD 163296, are classified as (pre-)transition disks with a central cavity (or large inner gap). While HD 163296 does not have the traditional (pre-)transition disk SED and only some evidence of possible inner clearing, it nonetheless has other structures in the disk and proposed companions, similar to the remaining targets in the sample and was therefore included here \citep{Espaillat2014,Isella2016}. In addition to central cavities, sub-millimetre dust emission and near-IR scattered light imaging have revealed features such as rings, clumps, and spirals in all the disks. 
At distances of 60-160 pc, the extended disks of the targets are expected to be large enough to be resolved with the Very Large Telescope (VLT) at Paranal in the 8-12\,$\mu$m wavelength range. 

Below, we provide an overview of the structure and possible companions of the targets, specifically those inferred through direct imaging.

\subsection{HD 100546}
This disk is divided into an inner disk and an outer disk, separated by a single gap from \textasciitilde1-21\,AU \citep[e.g.][]{Bouwman2003, Grady2005,Menu_2015,Jamialahmadi2018, 2019ApJ...871...48P}. It is possible that the inner and outer disks are misaligned \citep{2019ApJ...871...48P,Kluska2020}. The outer disk has spiral structures that have so far only been detected in the near-IR \citep{Follette2017,Quillen2006} and there is a tentative detection of a bar-like structure across the gap which could indicate small-scale inflow or be the base of a jet \citep{Mendigutia2017,Schneider2020}. There have been some suggestions of warping in the inner and outer disk, but this has so far remained inconclusive \citep[e.g.][]{Quillen2006,Panic2014,Pineda2014,Walsh2017,Sissa2018,Kluska2020}.

There has been much discussion about possible companions. One companion, HD 100546 b, was identified at a separation of 55\,AU at a position angle of 9$^\circ$ \citep{Quanz2013,Currie2014,Quanz2015}. However, this has been called into question by \cite{Rameau2017}, who failed to detect any accretion at the planet location in H$\alpha$ and posit the L$'$ band ($3.8\,\mu$m) detection might be related to the chosen method of data reduction. The lack of detection in H$\alpha$ is supported by \cite{Cugno2019}. A different companion, HD 100546 c, may have been detected just inside the central cavity at \textasciitilde13\,AU \citep{Brittain2014,Currie2015}, although this too has been contested \citep{Fedele2015,Follette2017,Sissa2018}. ALMA observations at 1.3 mm have revealed a 6$\sigma$ point source of $92 \pm 9\,\mu$Jy at a position angle of 37$^{\circ}$ and a projected separation of 7.8\,AU, which could represent an additional planetary candidate \citep[herafter HD 100546 d;][]{Perez2020}. A final planet candidate has also been suggested by the presence of a Doppler flip observed in the disk $^{12}$CO kinematics. Such a planet would be embedded within the disk continuum emission region exterior to the gap, corresponding to a projected radial distance of $20.5 \pm 5$\,AU \citep{Casassus2019}.

\subsection{HD 163296}
Near-IR and sub-millimeter wavelength observations show that HD 163296 has four gaps. They are centred on 10\,AU, 50\,AU, 81\,AU, and 142\,AU with bright rings in between \citep[e.g.][]{Garufi2014,Isella2016,Isella2018}. 

Companions have been suggested based on their possible role in forming the ring structures in the disk. For example, \citet{Liu2018} fitted three half-Jovian-mass planets and \citet{Teague2018} found the radial pressure gradients can be explained by two Jupiter-mass planet at 83 and 137\,AU (see also \citealt{Teague2019}). Additionally, \citet{Pinte2018} found a Jupiter-mass companion at 223\,AU based on deviations from Keplerian velocity in the gas of the disk. So far, observations have not been able to confirm or rule out such companions due to a lack of sensitivity. \citet{Guidi2018} claim to have found a 5-6\,M$_{\mathrm{Jup}}$ companion at a separation of 50\,AU from the star in the L$'$ band with Keck/NIRC2, but neither this object nor the one proposed by \citet{Pinte2018} was found by \citet{Mesa2019}, who set upper limits of 3-5\,$\mathrm{M_{Jup}}$ on possible companions in the gaps of the disk with SPHERE H band (1.6\,$\mu$m) and K band (2.2\,$\mu$m) data. Due to extinction from the disk setting, these kinds of mass limits remain challenging, especially outside the gaps, as only a fraction of the intrinsic, modelled flux of the companion may be observable.

\subsection{HD 169142}
The disk around HD 169142 has been imaged at near-IR and at sub-millimetre wavelengths. Various teams have imaged two \citep{Fedele2017,Quanz2013,Momose2015,Pohl2017}, three \citep{Macias2017,Osorio2014}, or four \citep{Macias2017,Perez2019} rings around the star. The inner ring is located at ~20\,AU and is more than twice as bright as the outer rings. As a result, it was found in all the previously mentioned works. The three outer rings (located between 45\,AU and 80\,AU) are faint and close together, leading to blending in some observations and resulting in the different numbers of rings found in different studies.

Four disk features that could be associated with forming planets have been found. The first was found between the 20\,AU and 50\,AU dust rings by \citet{Osorio2014} at 7\,mm, the second was found in the L$'$ band just within the edge of the inner gap by \citet{Reggiani2014} and \citet{Biller2014}. However, the L$'$ band source was not recovered by either team in the J (1.3\,$\mu$m), H, or K bands and it is concluded by \citet{Biller2014} that the feature cannot be due to planet photospheric emission and must be a disk feature heated by an unknown source, although \citet{Reggiani2014} argue that the accretion of material in the gap enhances the L$'$ band flux, resulting in a lower mass planet, which is not as easily observed in other bands. The presence of circumstellar material with entrained dust grains spreading across the gap or being accreted onto a planet could also subject the planet to further extinction in the J band. \citet{Biller2014} detected the third source in the H band, with no L$'$ band counterpart, but \citet{Ligi2018} show that this is actually part of the inner ring. They did find another H band structure close to the star that is consistent with the detections by \citet{Biller2014} and \citet{Reggiani2014}, but it appears to be extended and they cannot rule out that it is not part of a marginally detected ring at the same separation. Finally, \citet{Gratton2019} combined different SPHERE datasets and suggest that this source could actually be a combination of two extended blobs observed in the disk. They find a different, fourth, feature located between the inner and outer rings that does not correspond to any of the previous detections and could indicate the presence of a $2.2 \pm 1.4\,\mathrm{M_{Jup}}$ planet.

\subsection{TW Hydra}
TW Hydra is a 3-15\,Myr old  T Tauri star \citep{Vacca2011,Weinberger2013,Herczeg2014}. At a distance of $60.14 \pm 0.06$ pc \citep{Gaia2020}, it is one of the nearest known hosts of a protoplanetary disk. Studies in the near-IR and sub-millimetre wavelength regimes have found between three and six different gaps in eight different locations between 0.6\,AU and 90\,AU \citep{Nomura2016,Tsukagoshi2016, Andrews2016, vanBoekel2017,Huang2018}.

\citet{Tsukagoshi2016} suggest the presence of a $\lesssim 26\,\mathrm{M_{\oplus}}$ planet interacting gravitationally with the gap at 22\,AU.  \citet{Tsukagoshi2019}  found an azimuthally elongated 1.3\,mm continuum source in the south-west of the disk at a radial separation of 54\,AU that could be either dust that has accumulated into a clump in a vortex or a circumplanetary disk associated with an accreting Neptune mass planet. \citet{Nayakshin2020} argue the feature can be explained by a Neptune-mass planet disrupted in the process of accretion and expelling dust into the circumstellar disk.  Observations with SPHERE suggest from the gap profiles that if planets are responsible for forming the gaps in the circumstellar disk, they are at most several 10\,$\mathrm{M_{\oplus}}$ \citep{vanBoekel2017}.

\subsection{HD 100453}
HD 100453 has been found to possess a misaligned inner disk, a gap between 1\,AU and 21\,AU, and an outer disk with two shadows, two spiral arms around 30\,AU, and a faint feature in the south-west of the disk \citep{Benisty2017,Kluska2020}. It also has an M dwarf companion at a separation of 125\,AU whose orbit is not aligned with the disk plane \citep{vanderPlas2019}.

Dynamical modelling has shown that tidal interactions with the M dwarf companion are responsible for at least some of the disk features, such as the spirals and the truncation of the outer disk \citep{Wagner2018,vanderPlas2019,Gonzalez2020}. However, they have also suggested that the presence of a planet is required to fully explain the origin of the features in the disk, particularly the misalignment between the inner and the outer disks \citep[e.g.][]{Nealon2020}. There have been no direct detections of planet candidates to date.

\begin{table*}[ht]
    \centering
    \caption{\label{tab:observations}Overview of the observations used in this paper. HD 100546 was observed as part of different programmes than the other observations, leading to the difference in filters and observation times. }
   \begin{tabular}{lllllll}
    \hline \hline
    Target & Instrument & Date & Filter & $\lambda_0 (\mu m)$ & $\Delta\lambda (\mu m)$ & Integration time (s)\\ \hline
    HD 100546 & VISIR & 28-04-2018 & J8.9 & 8.70 & 0.74 & 3600 \\
    & NEAR & 11-12-2019 & PAH1 & 8.58 & 0.41 & 540 \\
    & & & ARIII & 8.98 & 0.14 & 540 \\
    &&& PAH2 & 11.24 & 0.54 & 540 \\
    && 12-12 2019 & PAH2\_2 & 11.68 & 0.37 & 540 \\ \hline
    HD 163296 & NEAR & 14-09-2019 & PAH1 & 8.58 & 0.41 & 600\\
    &&13-09-2019 & NEAR & 11.25 & 2.5 & 600\\ \hline
    HD 169142 & NEAR & 13-09-2019 & PAH1 & 8.58 & 0.41 & 600\\
    &&& NEAR & 11.25 & 2.5 & 600\\ \hline
    TW Hya & NEAR & 13-12-2019 & PAH1 & 8.58 & 0.41 & 600\\
    && 16-12-2019 & NEAR & 11.25 & 2.5 & 600\\ \hline
    HD 100453 & NEAR & 12-12-2019 & PAH1 & 8.58 & 0.41 & 600\\
    &&& NEAR & 11.25 & 2.5 & 600\\ \hline
    HD 36112/MWC 758 & NEAR & 18-12-2019 & NEAR & 11.25 & 2.5 & 600\\ \hline
    \end{tabular}
\end{table*}

\subsection{HD 36112}
HD 36112 (MWC 758) has a large cavity with a radius of 32\,AU. Its broad outer disk has rings, clumps, and spiral arms \cite[e.g.][]{Dong2018,Wagner2019}.

For the spiral structures in the disk of HD 36112 to be caused by a perturber, it is estimated that it must have a mass of $\sim$5-10\,$\mathrm{M_{Jup}}$ \citep{Grady2013,Dong2015b}. However, upper limits on companion fluxes obtained in the same works and by \citet{Reggiani2018} rule out the presence of >$5\,\mathrm{M_{Jup}}$ planets beyond 0.6\,$\arcsec$, or 94\,AU. \citet{Reggiani2018} found an L band ($3.5\,\mu$m) point source at 18\,AU that they interpret as a planet with a circumplanetary disk that is embedded in the disk. \citet{Wagner2019} did not find this object in the L$'$ and M$'$ bands, even though they achieved better sensitivities. Instead, they found a point source at the outer end of one of the spiral arms that could be a planet with a CPD and could be responsible for driving the spirals.

\section{Observations and data analysis}
\label{sec:obs}
Observations of HD 100546 were obtained during April 2018, with the VLT Imager and Spectrometer for the mid-IR (VISIR, \citealt{Lagage2004}), and of all six disks during the science verification of its upgrade, with NEAR (\citealt{Kasper2017}) in September and December of 2019. The benefit of NEAR is its use of AO, which results in improved angular resolution, PSF stability, and sensitivities (a factor of $\sim$4) across the N-band. An overview of the observations used in this paper is presented in Table \ref{tab:observations}. 

For all targets, all observations were taken in the pupil tracking mode, where the derotator is turned off to allow for field rotation during the observation sequence. For the NEAR observations, AO was enabled and the targets themselves were used as the reference star for wavefront sensing. The chopping and nodding sequence was enabled to subtract the sky background. In the VISIR data, the chop throw is 8$\arcsec$ in the direction perpendicular to the nodding direction; whereas, in the NEAR data, the chop throw is 4.5$\arcsec$ in the parallel direction. Since the throw determines the useful field of view, the VISIR and NEAR\ data have an effective field of view of 16$\arcsec$x16$\arcsec$ and 9$\arcsec$x9$\arcsec$, respectively. The VISIR data have a chopping frequency of 4\,Hz and a detector integration time (DIT) of 0.012\,s. The NEAR data have a chopping frequency of 8\,Hz and a DIT of 0.006\,s. Both NEAR and VISIR have platescales of 0.0453$\arcsec$.  

The standard VISIR data reduction pipeline\footnote{\url{https://www.eso.org/sci/software/pipelines/visir/visir-pipe-recipes.html}} is not suited to reduce data taken in the pupil tracking mode, so special purpose python scripts were employed to reduce and analyse the data. VISIR and NEAR data are delivered in chop difference images with integration times of 20-50\,s each. Data from the different nod positions are subtracted from each other and the resulting images are derotated. The beams from the chopping and nodding from all images are then median combined with 3$\sigma$ sigma clipping into a single master image. Only the VISIR observations of HD 100546 have a reliable reference star (HD 93813) with which to calibrate the result, leading to an observed flux of $27 \pm 3$\,Jy. For HD 100546 observations in other bands and in the cases of HD 163296, HD 169142, and TW Hya, we used the flux predicted by the \textsc{ProDiMo} models (described in Section \ref{section:prodimo}) to calibrate the data. Since the model is fitted to SED data from a collection of previous observations of the targets taken with other instruments, including data around 8-12\,$\mathrm{\mu m}$, it is the most accurate way available to determine the brightness in the images and this allowed us to calculate the flux in the specific wavelength ranges of the different filters \citep{CDAG4,Woitke2019}. The calibration is done by multiplying the model fluxes with the filter and sky transmissions and averaging the total flux over the required wavelength range. This is then set as the total flux of the data. As there are no models available for HD 100453 and HD 36112, the averages of previous flux measurements in similar filters had to be used \citep{vanBoekel2005,Carmona2008,Verhoeff2009,Marinas2011,Khalafinejad2016,Li2018}. 

\begin{figure*}
    \begin{minipage}{0.45\textwidth}
    \includegraphics[scale=1.]{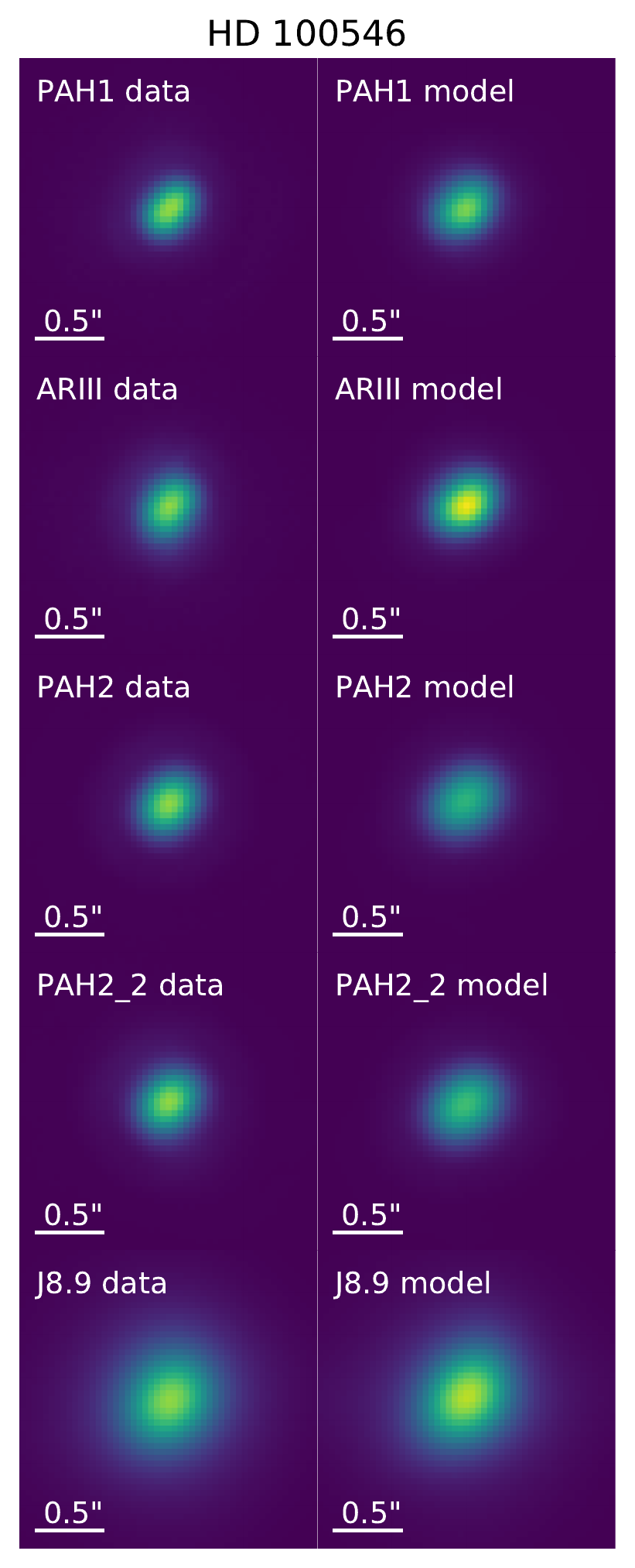}
    \caption{Master images (left) and model images (right) of HD 100546 in various filters. North is up and east is left in all images. The observations were scaled to have the same flux as the model images. The PAH1, ARIII, PAH2, and PAH2\_2 filter master images were taken with NEAR and show a resolved, inclined disk. The J8.9 data were taken with VISIR and are more extended compared to the NEAR data due to image elongation from the telescope resulting in a distorted and enlarged PSF. The model images provide a good match for the master images in each filter. }
    \label{fig:hd100456}
    \end{minipage}\hfill
    \begin{minipage}{0.45\textwidth}
    \includegraphics[scale=1.]{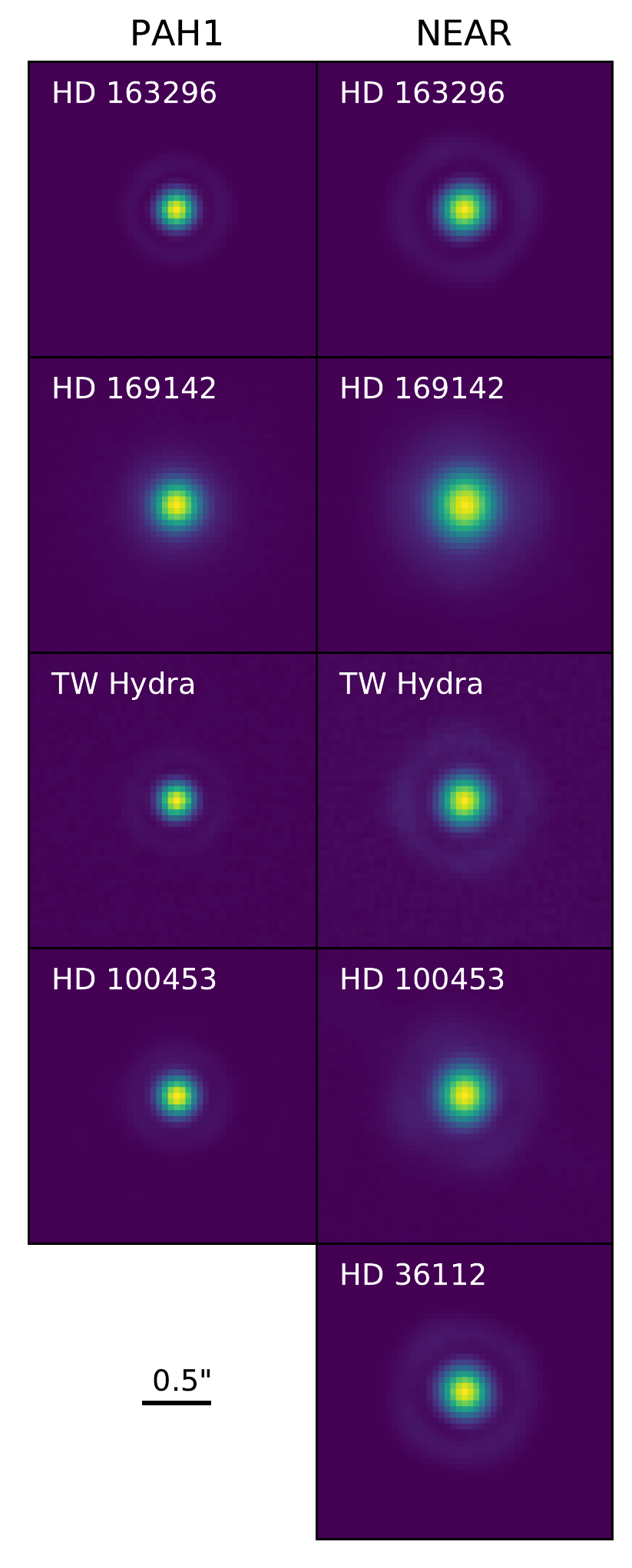}
    \caption{Normalised master images of the disks observed with NEAR. North is up and east is left in all images and the scale bar in the bottom left indicates 0.5$\arcsec$. The left column shows the disks in the PAH1 filter and the right column in the NEAR filter. HD 163296 and TW Hydra are unresolved in both filters. HD 36112 was not imaged in the PAH1 filter, but it is unresolved in the NEAR filter. Compared to these images, it can be seen that HD 169142 and HD 100453 are more extended in both filters. }
    \label{fig:disks}
    \end{minipage}
\end{figure*}

The final master images of the disks are shown in Figures \ref{fig:hd100456} and \ref{fig:disks}. The star is not visible in any of the images as it does not contribute significantly to the flux in the mid-IR (<10\% of the total flux in the \textsc{ProDiMo} models). Most of the central emission at these wavelengths is from unresolved inner disks or inner rims of outer disks.

\subsection{HD 100546}
The master images of HD 100546 in the different filters are shown in Figure \ref{fig:hd100456}, along with the corresponding model images after convolution with an appropriate PSF. For the J8.9 filter, this is the PSF of reference star HD 93813. While there were no appropriate flux calibration observations for the other filters, point sources were observed in the PAH1 and ARIII filters, which were used as PSF references. For the PAH1 filter and the ARIII filter, we used our own observations of HD 163296 and HD 27639, respectively. As there were no reference PSFs available in either the PAH2 or PAH2\_2 filters, we used scaled versions of the ARIII reference instead. Since the different filters on the NEAR instrument result in similar sensitivities over time, and the observations in the different filters have similar exposure times, all master images are expected to have similar sensitivities. The exception are the observations with the J8.9 filter which were taken with VISIR and where the increased observation time compensates for the lack of AO, meaning the final sensitivity of the master image is still expected to be similar to those in the other filters. While the disk is resolved in all filters, the VISIR data are clearly more extended than the NEAR data. The J8.9 band contains both the PAH1 and ARIII bands and so the VISIR image would be expected to have a similar extent as the NEAR images in these bands. Some of the difference is because the AO on NEAR means the images are more compact, but mostly due to the telescope operations during the VISIR observations. During this night, there was a decrease in the precision of the altitude axis of the telescope, resulting in elongation of the image along the paralactic angle \citep{deWit2020}. As this was at an angle of 40 degrees with the semi-major axis of the disk, the image is smeared along both axes and the smearing is not immediately obvious without a comparison. This is accounted for by using a reference PSF of the standard star HD 93813. Since this data set was taken immediately preceding the science observations in the same filter, it has a similar smearing effect.

The central bright emission in each image is from the unresolved inner disk, as the star is expected to be an order of magnitude fainter than the disk at mid-IR wavelengths based on the model data. Beyond that, emission is expected to be dominated by the inner rim of the outer disk, which is irradiated by the star and puffed up as a result. The rest of the outer disk is not warm enough to be detected in the image. 

Using a Levenberg-Marquardt algorithm and least squares statistic to fit a simple two dimensional Gaussian to the surface brightness of the disk in each filter results in an average position angle of $141 \pm 2^\circ$. Since we are fitting a two-dimensional function to a three-dimensional disk, we are sensitive to projection effects. This is especially the case because the inner wall of the outer disk is only visible on the far side of the disk and not on the close side. This means what we are calculating is actually the position angle of the two-dimensional projection of the disk, which we call the projected position angle. We also applied this method to model images of HD 100546 at the same wavelengths and found that the projected position angle is $\sim130^\circ$, compared to the input of $140^\circ$, so we expect a difference between the projected position angle and the real position angle of roughly 10 degrees. This would still be in agreement with previous position angle values of 135-150$^\circ$ \citep{Miley2019,Casassus2019, Jamialahmadi2018,Mendigutia2017,Pineda2014,Avenhaus2014,Walsh2014,Leinert2004}. A more precise determination of the disk orientation requires extensive modelling and is outside the scope of this paper. 

The deprojected disk has a full-width-half-maximum (FWHM) of 0.82$\arcsec$ in the J8.9 filter and 0.35$\arcsec$-0.41$\arcsec$ in the other filters. The larger size of the J8.9 image is due to the above-mentioned PSF smearing from uncertainty in the altitude axis of the telescope. The FWHM values for all the disks and filters are listed in Table \ref{tab:fwhm}. From the disk FWHM and the PSF FWHM (the diffraction limit is $0.22\arcsec-0.30\arcsec$ depending on the filter), we can calculate the true size of the emitting region, assuming that both the data and the PSF are well described by Gaussian functions \citep[e.g.][]{Marinas2011,vanBoekel2004}, as follows: 
\begin{equation}
    \mathrm{FWHM_{disk}} = \sqrt{\mathrm{FWHM_{data}}^2-\mathrm{FWHM_{PSF}}^2}
    \label{eq:fwhm}
.\end{equation}
Due to the PSF smearing in the J8.9 image, we used the reference PSF FWHM rather than the theoretical diffraction limit for this filter. Since the other data were observed with the NEAR instrument, which thanks to its adaptive optics is expected to have a Strehl ratio of close to one \citep{Kasper2017}, the FWHM of a point source PSF corresponds to the diffraction limit. This can be seen in the data of HD 163296, TW Hydra, and HD 36112, as is discussed in Sect. \ref{sec:other_sources}.
The deconvolved FWHM of all resolved sources and the corresponding 5$\sigma$ upper limits for unresolved sources are also listed in Table \ref{tab:fwhm}. While spectroscopic data show that the disk is more extended in PAH emission bands \citep{vanBoekel2004,Verhoeff2009}, the PAH1 and PAH2 filter images are no more extended than their continuum counterparts. This is because the extent of the emission is averaged over the filter wavelength range and the PAH emission is estimated to be around 22\% of the total flux in the PAH1 filter and 13\% in the PAH2 filter \citep{vanBoekel2004}. As a result, both PAH filter images are dominated by the continuum emission and have similarly sized emission regions as the images in the continuum filters. The 2$\sigma$ discrepancy between the J8.9 and the PAH1 and ARIII deconvolved FWHM means the errorbars on the J8.9 image are probably underestimated, possibly due to a worsening of the smearing effect as the night went on.

Removing the PSF component along both axes also gives a more accurate inclination, since the semi minor axis of the disk is relatively more extended by the PSF than the semi-major axis. The calculated inclination is $47 \pm 3^\circ$. The projection effect is not expected to be as strong here, since even on the model data the resulting inclination was well within $1\sigma$ of the input value. The projected inclination is in agreement with literature inclination values of 42-50$^\circ$. \citep{Miley2019,Casassus2019,Jamialahmadi2018,Mendigutia2017,Pineda2014,Avenhaus2014,Walsh2014}. This value is the combined inclination across all the available filters, except for J8.9 due to the deformed PSF in this image.

\subsection{Other sources} 
\label{sec:other_sources}
HD 163296 is unresolved in both filters and has FWHMs around the diffraction limit of the telescope which is 0.22$\arcsec$ in the PAH1 filter and 0.30$\arcsec$ in the NEAR filter. This results in 5$\sigma$ upper limits of 7\,AU and 6\,AU, respectively. Previous mid-IR observations between 8\,$\mathrm{\mu}$m and 13\,$\mathrm{\mu}$m have not resolved the disk, but set an upper limit on the FWHM of the emission region of 21\,AU at 11.7\,$\mathrm{\mu}$m \citep{Jayawardhana2001,vanBoekel2005,Marinas2011,Li2018}. Our images of HD 163296 improve on the emission size upper limits by a factor of three.

HD 169142 is the most resolved disk in the sample after HD 100546. The measured and deconvolved FWHM are listed in Table \ref{tab:fwhm}. Additionally, the measured projected inclination of the deconvolved disk is $13 \pm 2^\circ$, which is in agreement with previously measured inclinations of $13 \pm 1^\circ$ \citep{Perez2019,Panic2008,Raman2006}. 

TW Hydra is unresolved in our observations with upper limits of 3\,AU in the PAH1 band and 49\,AU in the NEAR band. The high limit in the NEAR band is due to the data being taken with the coronograph. While this allows for increased sensitivity for finding planets, it also means that the extent has to be calculated with the off-axis chop and nod beams. Based on the PAH1 data taken the same night, the beams are expected to be smeared by $\sim$\,10\%. These limits are consistent with previous interferometry measurements which found the size of the emitting region of the disk to be 1-2\,AU around 8-12\,$\mathrm{\mu}$m \citep{Ratzka2007,Arnold2012}. 

HD 100453 is resolved in both bands. Similar to TW Hydra, the NEAR band images of HD 100453 were taken with the coronograph, resulting in a 10\% error in the extent of the emission region. The difference between the deconvolved PAH1 and NEAR band sizes suggests this might still be an underestimate. The disk has a calculated projected inclination of $35 \pm 5^\circ$, which is in agreement with literature values of the inclination of 30-38$^\circ$ \citep{Rosotti2020,Long2017,Benisty2017,Wagner2015}. 

Finally, HD 36112 is unresolved, with a NEAR band upper limit of the size of the emission region of 13\,AU. This is an improvement by almost a factor of 10 over previous observations which set an upper limit of 120\,AU on the 11.7\,$\mathrm{\mu}$m emission size \citep{Marinas2011}.

\begin{table}
    \centering
    \caption{\label{tab:fwhm} FWHM of the disks in each filter is given in arcseconds. HD 100546 is clearly resolved in all bands. HD 169142 and HD 100453 are resolved in both the PAH and NEAR bands, while HD 163296, TW Hydra, and HD 36112 are unresolved point sources. For resolved images, the FWHM after deconvolution is listed in AU. For unresolved images, the 5$\sigma$ upper limits are listed instead.}
    \begin{tabular}{llcc}
         \hline \hline
         Object & Filter & FWHM$_\mathrm{data}$ ($\arcsec$) & FWHM$_\mathrm{disk}$ (AU)\\ \hline
         HD 100546 & J8.9 & 0.82 $\pm$ 0.10 & 61 $\pm$ 11  \\
          & PAH1 & 0.349 $\pm$ 0.003 & 28.9 $\pm$ 0.5\\
          & ARIII & 0.356 $\pm$ 0.002 & 29.0 $\pm$ 0.3\\
          & PAH2 & 0.392 $\pm$ 0.002 & 28.2 $\pm$ 0.4\\
          & PAH2\_2 & 0.414 $\pm$ 0.002 & 30.5 $\pm$ 0.3\\ \hline
         HD 163296 & PAH1 & 0.216 $\pm$ 0.002 & <7 \\
          & NEAR & 0.282 $\pm$ 0.001 & <6\\ \hline
         HD 169142 & PAH1 & 0.336 $\pm$ 0.003 & 28.6 $\pm$ 0.5\\
          & NEAR & 0.465 $\pm$ 0.003 & 41.1 $\pm$ 0.5\\ \hline
         TW Hydra & PAH1 & 0.219 $\pm$ 0.001 & <3\\. 
          & NEAR & 0.297 $\pm$ 0.029 & <49\\ \hline
         HD 100453 & PAH1 & 0.234 $\pm$ 0.002 & 9.3 $\pm$ 0.6\\
          & NEAR & 0.352 $\pm$ 0.035 & 20.7 $\pm$ 6.6\\ \hline
         HD 36112 & NEAR & 0.315 $\pm$ 0.002 & <13\\ \hline
    \end{tabular}
    
\end{table}

\section{Protoplanetary disk modelling with ProDiMo} \label{section:prodimo}

\begin{table}
    \centering
    \caption{DIANA SED-fit parameters for the HD 100546 system used in the \textsc{ProDiMo} disk model. Parameters that were modified to improve the mid-IR fit are included in parenthesis.}
   \begin{tabular}{lll}
    \hline \hline
        Parameter               & Symbol             & Value  \\ \hline
        Stellar Mass            & $M_*$              & 2.5 $M_{\rm \odot}$  \\
        Stellar Luminosity      & $L_*$              & 30.46 $L_{\rm \odot}$ \\
        
        Effective Temperature   & $T_{\rm eff}$      & 10470 K  \\
        Interstellar Extinction  & A$_{\rm V}$ & 0.22 mag  \\
        \hline

        Dust composition: \\
        \:  Mg$_{0.7}$Fe$_{0.3}$SiO$_3$ & & 58.17 $\%$ \\
        \:  Amorphous carbon & &  $16.83\%$\\
        \:  Vacuum &  &  $25\%$            \\
        Dust size power law & p & 3.34\\

        \hline

        Disk Inner Zone \\

        \:  Mass               & $M_{\rm d}$    & $8.81\times10^{-8}$  $M_{\rm \odot}$ \\
        \:  Inner Radius       & $R_{\rm in} $  & 0.55\,AU \\ 
        \:  Outer Radius       & $R_{\rm out} $ & 4.00\,AU \\
        \:  Col. Density Power Index        & $\epsilon$ & 0.35\\

        \: Minimum dust size & a$_{\rm min}$ & 0.042$\mu$m\\
        \: Maximum dust size & a$_{\rm max}$& 2.9 $\mu$m \\

        \:  PAH abundance       & $f_{\rm PAH} $  & 0.0028  \\ 

        \hline

        Disk Outer Zone \\

        \:  Mass               & $M_{\rm d}$    & $7.15\times10^{-3}$  $M_{\rm \odot}$ \\
        \:  Inner Radius       & $R_{\rm in} $  & 19.34 (22.3) \,AU \\ 
        \:  Outer Radius       & $R_{\rm out} $ & 600\,AU \\
        \:  Tapering Radius    & $R_{\rm tap} $ & 100\,AU \\
        \:  Col. Density Power Index   & $\epsilon$ & $1.12$ \\

        \: Minimum dust size & a$_{\rm min}$ & 0.042$\mu$m\\
        \: Maximum dust size & a$_{\rm max}$& 2983 $\mu$m \\
        \:  PAH abundance       & $f_{\rm PAH} $  & 0.0028 (0.0034)  \\ 

        \hline
        
        \: Inclination             & i                &  $42^{\circ}$ \\
        \: Dust to Gas Ratio       &  $d/g$           & 0.01 \\

        \hline
        
    \vspace{1ex}
    \end{tabular}
    
    \label{table:disk_params}
\end{table}



\begin{figure}
    \centering

    \includegraphics[width=\textwidth/2]{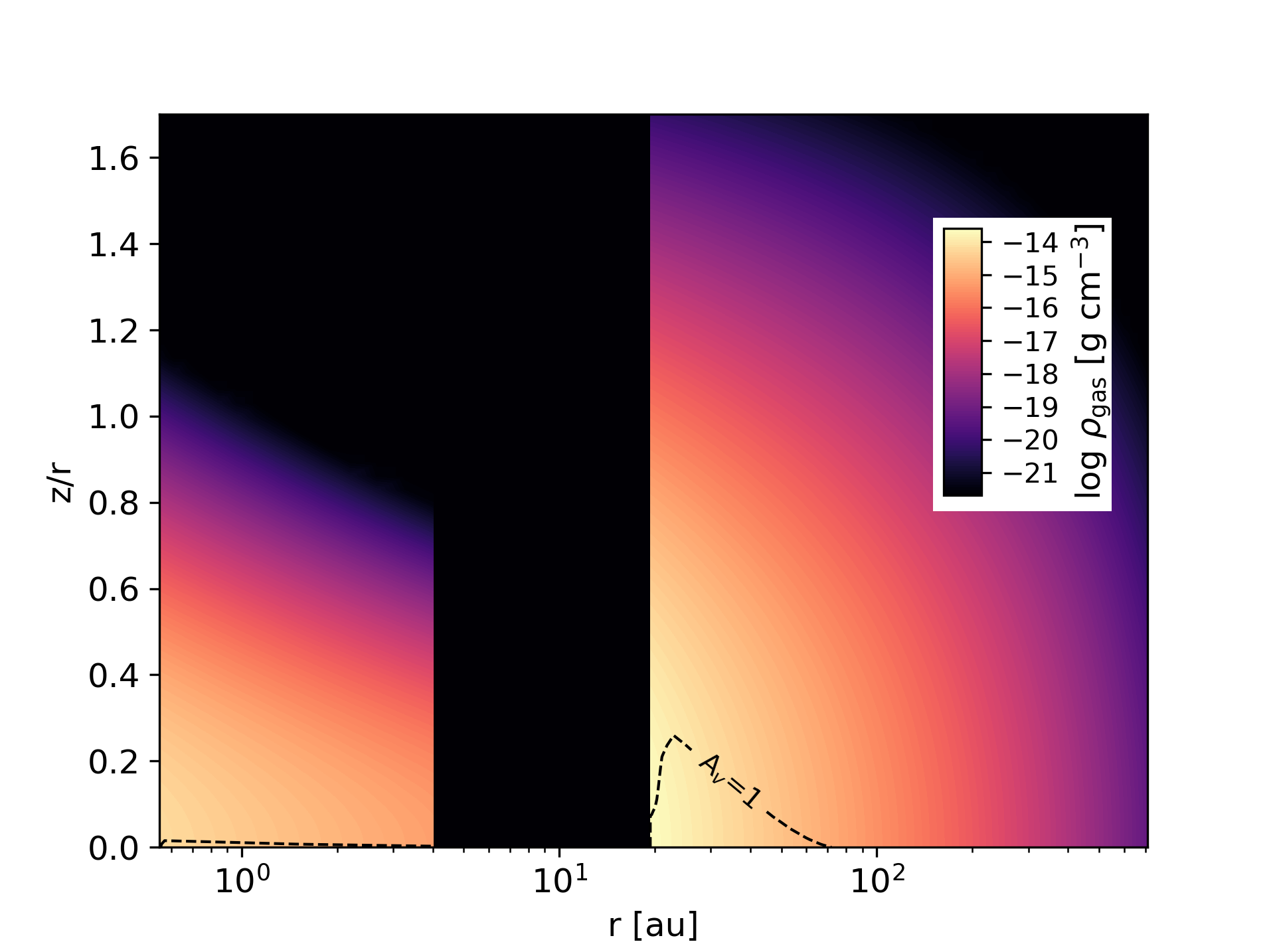}
    \caption{Gas density profile of the \textsc{ProDiMo} HD\,10056 disk model.  The dashed contour line traces the surface where the minimum optical extinction $A_{\rm V}$ in the combination of the vertical or radial direction is 1. } 
    \label{fig:nhtot}

    \centering
    \includegraphics[width=\textwidth/2]{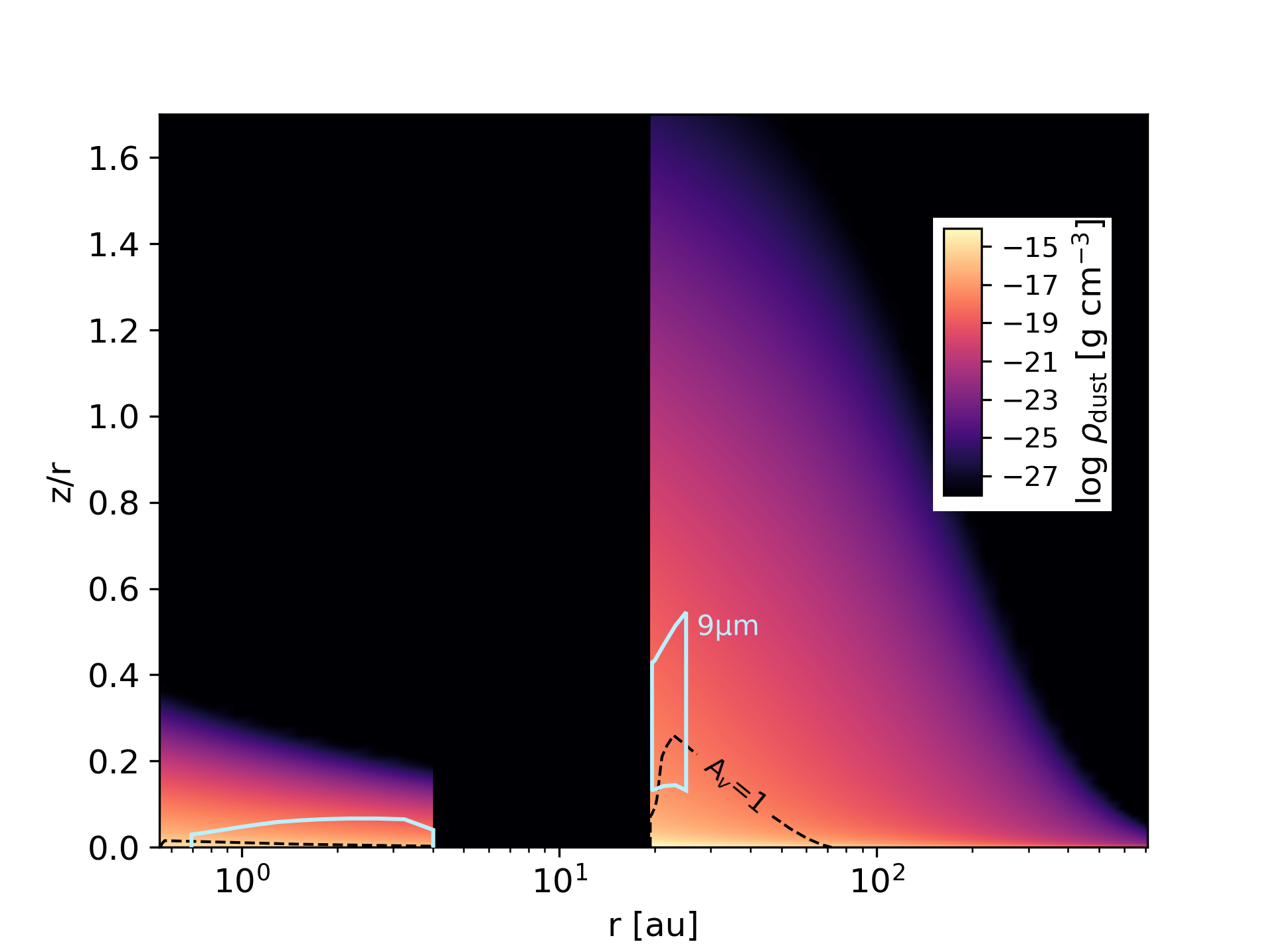}
    \caption{Dust density profile of the \textsc{ProDiMo} HD\,100546 disk model.  The light blue contour outlines the region where half of the total 9 $\mu$m emission originates.  The dashed contour line traces the surface where the minimum optical extinction $A_{\rm V}$ in the combination of the vertical or radial direction is 1. }
    \label{fig:rhod}

\end{figure}

We used the radiation thermo-chemical disk model \textsc{ProDiMo}\footnote{\url{https://www.astro.rug.nl/~prodimo/}} \citep{2009A&A...501..383W,2010A&A...510A..18K,2011MNRAS.412..711T} to simulate observations of the HD100546 system.  \textsc{ProDiMo}  self-consistently and iteratively determines the physical and chemical state anywhere within the disk with a frequency dependent 2D dust continuum radiative transfer, including gas-phase and photo-chemistry, ice formation, and non-LTE heating and cooling mechanisms. \textsc{ProDiMo} performs a 2D continuum radiative transfer with a ray-based, long-characteristic, accelerated $\Lambda$-iteration method at every disk grid point to calculate the local radiation field $J_{\nu}(r,z)$ \citep{2009A&A...501..383W}. The full radiative transfer methodology is described in \citet{2009A&A...501..383W}.  We adopt the standard DIANA\footnote{https://dianaproject.wp.st-andrews.ac.uk/data-results-downloads/fortran-package/} dust opacities as described in \citet{Woitke_2016} and \citet{Min_2016}.  

The parameters for the HD 100546 disk model were derived from the SED fitting work done as part of the European FP7 project DIANA\footnote{More information about the fitted stellar and disk parameters, the 2D modelling results, and the predicted observables can be found at \url{http://www-star.st-and.ac.uk/~pw31/DIANA/DIANAstandard} } \citep{Woitke2019}.  Parameters of the HD 100546 disk and stellar model can be found in Table \ref{table:disk_params} and the 2D gas density profile can be found in Fig. \ref{fig:nhtot}.  The fitting was performed for a pre-Gaia distance of 103\,pc \citep{Ancker1997}. Further details regarding the disk modelling and SED fitting process can be found in Appendix \ref{appendix:SED_fitting}.

As \textsc{ProDiMo} finds formal solutions to the continuum radiative transfer during the calculation of the SED, the resulting modelled intensity can be visualised as an image.  \textsc{ProDiMo} includes only the effect of isotropic scattering, and hence the preferential forward-scattering of light by larger dust grains is not represented realistically. As a result, the \textsc{ProDiMo} model appears brighter on the far side than on the near side and it cannot reproduce the observed asymmetry in brightness of actual disks.  While this effect is cancelled
out in the disk SED model and radial intensity profile, it must be taken into consideration when comparing the model image to data on a per-pixel basis.  The resulting \textsc{ProDiMo} data cube was attenuated by multiplying each synthetic disk image with the VISIR and NEAR relative filter transmission curves created with the VISIR imaging detector and VISIR calibration unit, and then by the sky transmission\footnote{\url{https://www.eso.org/observing/etc/bin/gen/form?INS.MODE=swspectr+INS.NAME=SKYCALC}} at each wavelength. Subsequently the data cube was flattened into a single image for each filter.  The images were then convolved with a reference PSF to simulate our observations. This was HD 93813 for the J8.9 filter, HD 27639 for the ARIII filter, and the HD 163296 data for the PAH1 and NEAR filters. For the PAH2 and PAH2\_2 filters, reference PSFs were not available and the PSF from the ARIII filter was scaled to the new central wavelength and used instead.

\section{Comparison to ProDiMo disk models}
\label{sec:comparison}

\subsection{Spectral energy distribution HD 100546}

\begin{figure}
    \centering
    \includegraphics[scale=0.6]{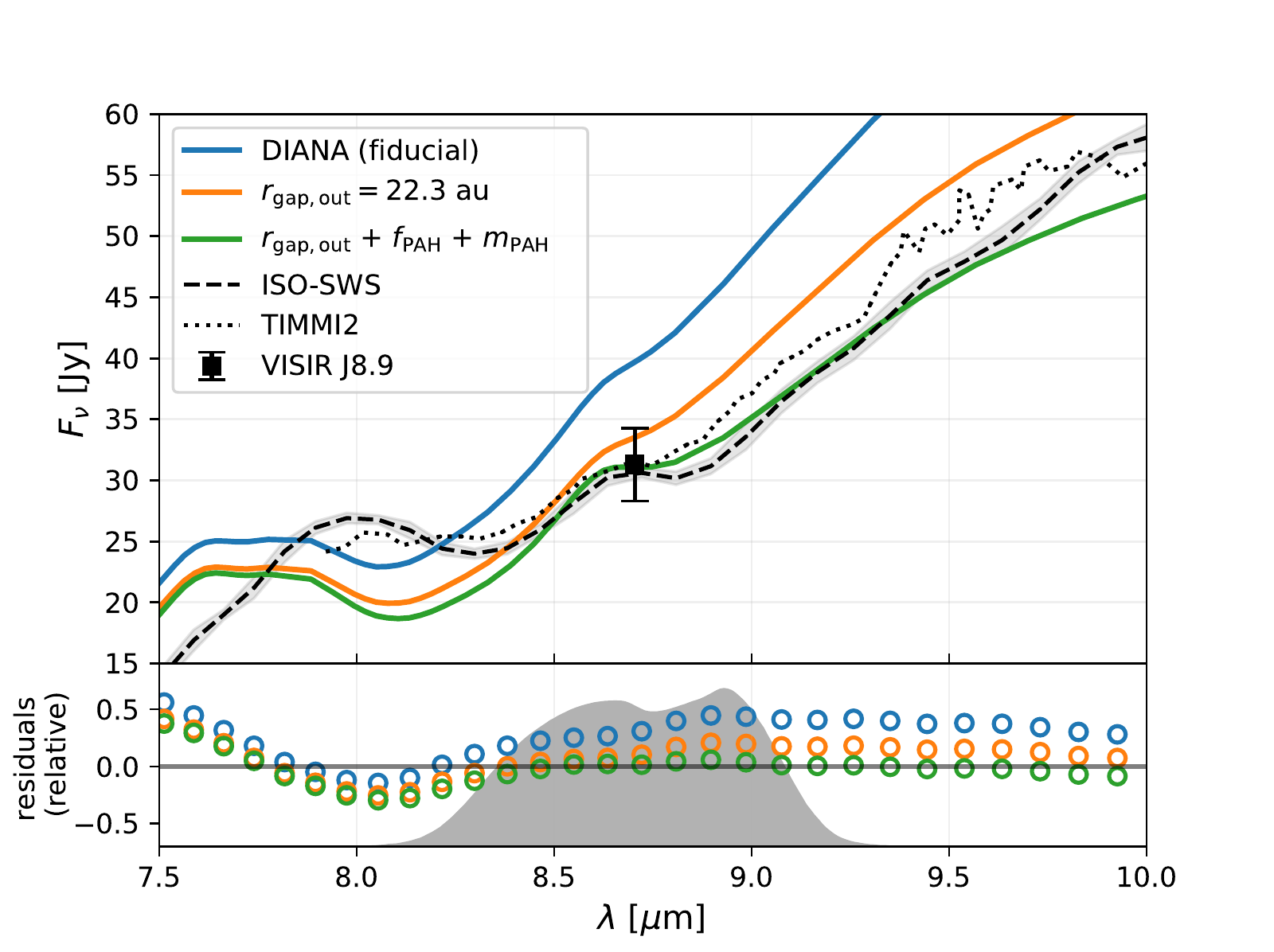}
    \caption{Comparison between the fiducial \textsc{ProDiMo} HD 100546 disk model \citep{Woitke2019} and multi-parameter variants of the model. We include the observational VISIR data corrected for sky transmission and additional observational data \citep{1998A&A...332L..25M,vanBoekel2004,2010A&A...514A...1I}.  The grey filled area illustrates the J8.9 filter response curve (arbitrary vertical scaling). Residuals between the various disk models and the \textit{ISO-SWS} spectrum are shown in the lower panel as the ratio between the model SED and the observed spectrum.} 
    \label{fig:models}
\end{figure}

Figure \ref{fig:models} illustrates the resulting SED for variants of the fiducial \textsc{ProDiMo} HD 100546 model between 7.5 and 10\,$\mathrm{\mu}$m, along with the averaged flux of the J8.9 band observation. The VISIR observations are included in black, as are the flux measured by \emph{AKARI} and the spectrum from \emph{ISO} \citep{1998A&A...332L..25M, 2010A&A...514A...1I}.  Near 8.7\,$\mu$m, the observational data to which the SED was fit includes the \textit{ISO-SWS} spectrum and a photometric data point from \textit{AKARI} with the S9W filter \citep{1998A&A...332L..25M, 2010A&A...514A...1I}.  While our data are in agreement with previous observational data, the expected flux of the basic \textsc{ProDiMo} model falls outside the uncertainty interval.   We consider both disk parameter modifications included and not included in the previously performed SED fitting process that may improve upon the local fit in the mid-IR without reducing the quality of the global fit.

In our disk model, the continuum flux at 8.7\,$\mathrm{\mu}$m is emitted largely from the surface of the inner disk between 1-4\,AU,  while in the outer disk the 8.7\,$\mathrm{\mu}$m flux originates largely from the gap wall which is directly illuminated by the star and heated to $\sim$300\,K (see Fig. \ref{fig:rhod}).   Modifying the location of the cavity's outer rim ($r_{\in}$ of the disk outer zone) allowed us to reduce the temperature of the gap wall and reduce the continuum emission in the mid-IR.  We find the optimal balance between moving the gap outer wall further outwards and maintaining the quality of the global fit occurs where the gap wall is moved outwards from 19 to 22.3\,AU.  As demonstrated in Fig. \ref{fig:models} by the line $r_{in} = 22.3$\,AU, this brings the SED within formal agreement to our observed mid-IR flux. Of the observed excess flux over the continuum around 10\,$\mu$m, $\sim$\,60\% has been explained by the presence of amorphous olivine and crystalline forsterite emission features with the remainder explained by PAHs \citep{1998A&A...332L..25M}.   We thus also consider further refinements to the mid-IR fit by exploring the properties of the disk PAH population. These considerations can be found in Appendix \ref{appendix:PAH_fitting}.

 Across the wavelength coverage of the \textit{ISO-SWS} spectrum, we reduced the sum of the squares of the ratio between the old fit $F_{\nu}^{\rm old}$ and the new fit $F_{\nu}^{\rm new}$, that is $\Sigma (F_{\nu}^{\rm new}/F_{\nu}^{\rm old})^2$, from 12.6 to 4.2. It should be noted that while dust settling allows for a variety of average grain sizes across the vertical extent of the disk model,  dust grains are not radially segregated by size in \textsc{ProDiMo}, such that within our model's disk zones, every grid column contains the same underlying dust grain size distribution.  Hence we can solve for only one gap outer radius, rather than a radius for each corresponding grain size.

\subsection{Radial intensity profile}
Radial intensity profiles of all the disks in the sample in the different filters were constructed by azimuthally averaging over the deprojected disks for both the observations and the convolved models and this is shown in Figs. \ref{fig:radial} and \ref{fig:radial2}. In all cases the radial profile is dominated by the telescope PSF. The unresolved sources show clear Airy rings in the images (see Fig. \ref{fig:disks}). The Airy rings are less obvious in the resolved sources and the central disk of the Airy pattern is larger, but they are still visible in the radial profiles. None of the profiles show signs of spirals, rings, or other features in the extended disk. Although the models do not include these previously observed features, this result is still consistent with the models, which show that the mid-IR emission is dominated by the central regions and the outer regions where features have been detected at other wavelengths contribute less than 5\% of the flux at 8.7\,$\mathrm{\mu}$m. 

For most models used in this comparison, the distance was measured before the Gaia data release. With the release of the Gaia data, it appears that these distances were off by around 10\% in most cases (HD 100546, HD 163296, TW Hydra). For these disks, it was not necessary to rerun the model, as the differences between the old and new distances are small. Simply rescaling the model to the new distance is sufficient to compare the extent of the disks. However, for HD 169142, the difference between the distance assumed in the model and the distance measured by Gaia is more significant: The assumed distance is almost 30\% too large. Because of this, the model was rerun with an adapted luminosity for the new distance.

\subsubsection{HD 100546}

\begin{figure}
    \centering
    \includegraphics[scale=.8]{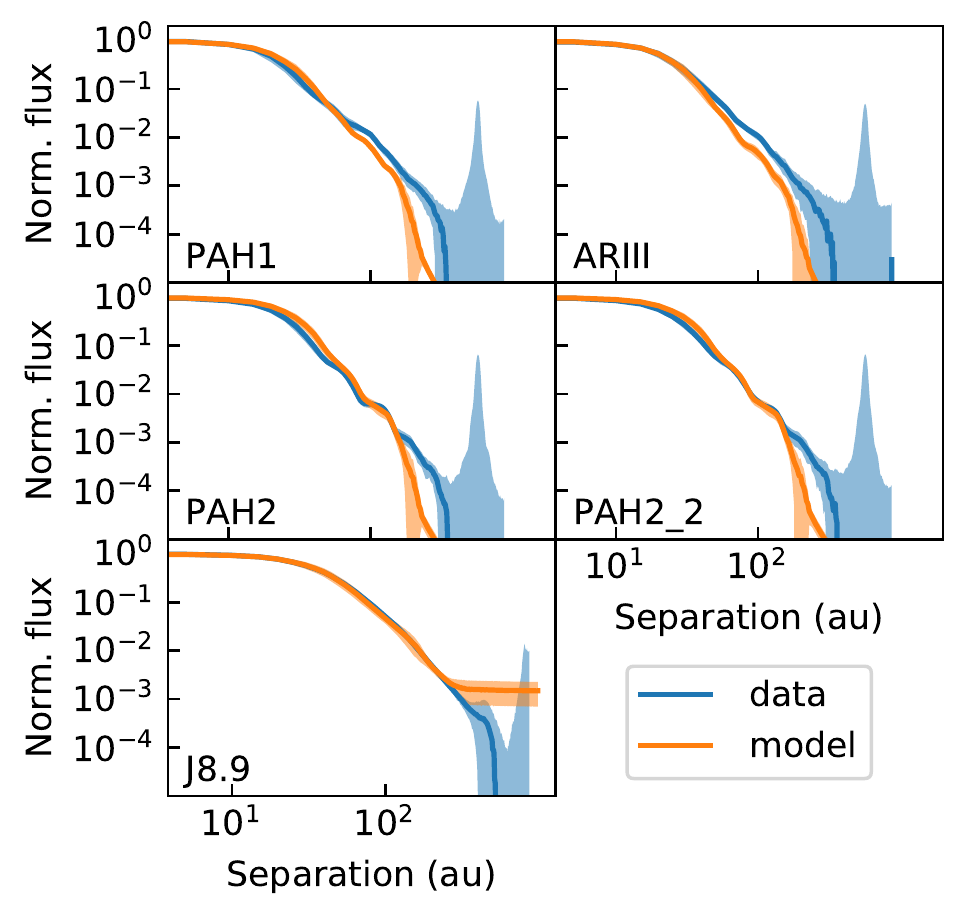}
    \caption{Radial profile of the HD 100546 protoplanetary disk in the PAH1, ARIII, PAH2, PAH2\_2, and J8.9 filters. The profile from the data is shown in blue with the 1$\sigma$ range in light blue. The profile from synthetic observations based on the \textsc{ProDiMo} model is shown in orange with the 1$\sigma$ range indicated in lighter orange. }
    \label{fig:radial}
\end{figure}

\begin{figure}
    \includegraphics[scale=.65]{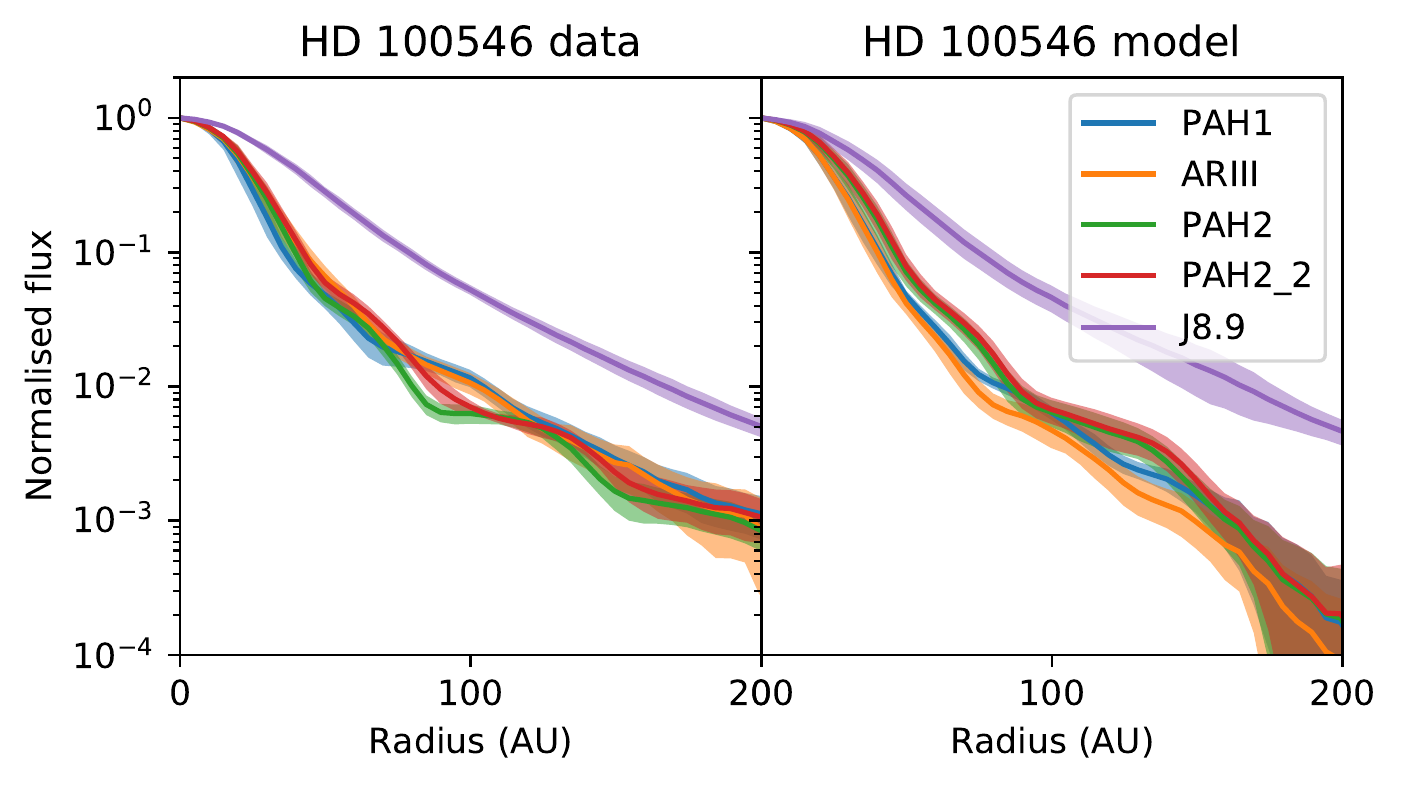}
    \caption{Radial flux profile of the HD 100546 protoplanetary disk in the PAH1, ARIII, PAH2, PAH2\_2, and J8.9 filters, with the real data profiles on the left and the synthetic data profiles on the right. The shaded areas indicate 1$\sigma$ errors for the data and confidence intervals for the models. For the model profiles, these intervals come from the PSF convolution and the azimuthal averaging and deprojection. In both the data and the model, it can be seen that the radial extent at 1/10$^{th}$ the maximum flux is smaller for the shorter wavelength filters (PAH1, ARIII) than for the larger wavelength filters (PAH2, PAH2\_2). This is expected as the PSF is larger for larger wavelengths. The J8.9 data, both real and synthetic, remain far more extended due to the smeared PSF.}
    \label{fig:radial2}
\end{figure}

\begin{figure}
    \begin{minipage}{0.48\textwidth}
    \includegraphics[scale=.6]{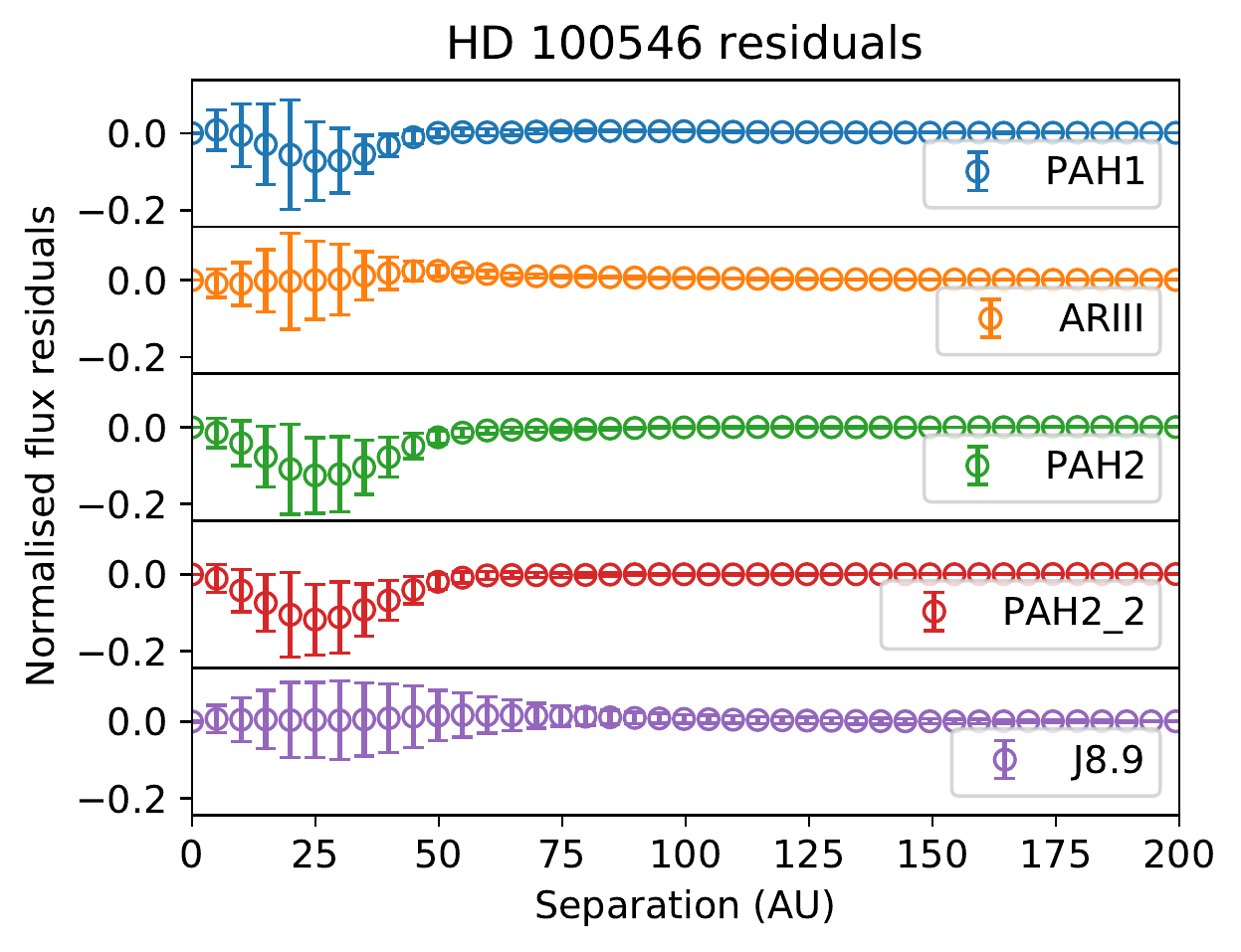}
    \caption{\label{fig:residuals} Residuals from subtracting the radial profile of the synthetic data from that of the observed data in each of the observed filters. The errorbars indicate 1$\sigma$ uncertainties. The residuals all being within 1$\sigma$ of 0 show that the model represents the data well. }
    \end{minipage}
\end{figure}

The normalised radial flux distribution of both the real, deprojected data in each filter and the corresponding simulated data are shown in Fig. \ref{fig:radial}. The model and the data are in good agreement out to $\sim$160\,AU, where the noise starts to dominate the signal. The peak in the noise in the data is caused by the source subtraction in the chopping and nodding. The subtraction shadows are located at $\sim$500\,AU (4.5$\arcsec$) in the four NEAR filters and at $\sim$900\,AU (8$\arcsec$) in the J8.9 filter. 
In Fig. \ref{fig:radial2} we can compare the different filters to each other for the observed and synthetic data. In both cases the shorter wavelength filters PAH1 and ARIII result in narrower profiles with a smaller FWHM than the longer wavelength filters PAH2 and PAH2\_2. Due to the smearing of the PSF, the J8.9 filter profile is much wider in both cases. The residuals from subtracting the model curves from the data are shown in Fig. \ref{fig:residuals}. The errorbars in the image represent the 1$\sigma$ error. The residuals show that the synthetic data is a good representation of the real data. The residuals at larger separations are 0 because the chopping and nodding process removes the background emission from the data and the model does not include sky or instrument background emission. 

\subsubsection{HD 163296}
Previous observations in near-IR\ and sub-millimetre wavelengths show that HD 163296 has multiple bright rings \cite[e.g.][]{Garufi2014,Isella2016,Isella2018}. The \textsc{ProDiMo} model does not include rings, but instead assumes a flared, optically thick inner region up to 0.02$\arcsec$ and a shadowed outer region beyond that. As a result, it predicts that 95\% of the flux is contained within a radius of 0.01$\arcsec$ in the PAH1 band and within 0.04$\arcsec$ in the NEAR band. This makes the emitting region much smaller than in the case of HD 100546, where there is a cavity and the inner rim of the outer disk also contributes to the flux. It is also entirely consistent with an unresolved image.

\subsubsection{HD 169142}
ALMA observations have detected three bright rings between 0.2$\arcsec$ and 0.6$\arcsec$ ($45-80$\,AU) in the disk around HD 169142 \citep{Perez2019}. Again, the model does not include the rings, but instead divides the disk into an inner and an outer zone with a gap at 0.1$\arcsec$ (22\,AU), which is consistent with the inner gap seen at other wavelengths. 
Assuming the observed disk is described by a Gaussian function, the apparent size as defined by \textsc{ProDiMo} (the radius containing 95\% of the flux) corresponds to the 2$\sigma$ radius of the Gaussian, which is larger than the FWHM, which only contains half the flux. After deconvolution, the apparent size of HD 169142 is $24 \pm 1$\,AU in the PAH1 band and $35 \pm 1$\,AU in the NEAR band. This means that the inner gap is unresolved and part of the flux in both bands is from the inside of the inner ring, but the outer two rings are too faint to be observed. 

The HD 169142 model has an apparent size of 43 and 45\,AU in the PAH1 and NEAR bands. While this is approximately consistent with the observed apparent size in the NEAR band, there is a discrepancy with the smaller PAH band observation. This is consistent with observations by \citet{Okamoto2017}, who find that the size of the emitting region is much smaller at 8.6 and 8.8\,$\mathrm{\mu}$m than it is at 12.6\,$\mathrm{\mu}$m. They conclude that at wavelengths smaller than 9\,$\mathrm{\mu}$m, the inner disk and halo dominate; whereas, at wavelengths larger than 9\,$\mathrm{\mu}$m, the inner wall of the disk dominates which results in a larger observed size. Modelling performed by \citet{maaskant2014} suggests that gas flowing through disk gaps can contribute significantly to the observed ionised PAH emission. This could manifest as an increase in emission at $\sim$8\,$\mathrm{\mu}$m relative to $\sim$12\,$\mathrm{\mu}$m, corresponding to the angular size of a gap.  If the neutral PAH emission primarily originates from the gap wall, we would expect a correspondingly smaller emitting region for the predominantly $\sim$8\,$\mathrm{\mu}$m PAH flux. This difference is not reproduced by the model, leading to a mismatch with the data in the PAH band. This can be due to the complete lack of gas and dust in the model gap and hence lack of associated emission.

The previously derived inclination of $13 \pm 2^\circ$ is consistent with the model value of $13^\circ$. It is also consistent with previous literature \citep{Perez2019,Panic2008,Raman2006}.

\subsubsection{TW Hydra}
Studies in near-IR and sub-millimetre have found six gaps located between 0.11$\arcsec$ and 0.84$\arcsec$ ($6-44$\,AU) \citep{Tsukagoshi2016, Andrews2016, vanBoekel2017}. 
The model assumes an optically thin inner region corresponding to the inner gap and a dense outer region for the rest of the disk. All the emission in both bands is predicted to be from this thin inner region and the inner wall of the outer disk. The other gaps are not expected to be visible as they are further out in the disk, where there is no more emission. This means that there is an apparent size of 3-4\,AU in both filter bands and this is consistent with the observations being unresolved. More recent observations also suggest the central optically thin region may be much smaller than in the model, which would shrink the expected apparent size \citep[e.g.][]{vanBoekel2017,Andrews2016}. 

\subsubsection{HD 100453}
Deconvolving the data results in apparent sizes of $7 \pm 1$ and $18 \pm 1$\,AU in the PAH1 and NEAR filter bands, respectively. The contribution of PAHs to the flux in the PAH1 band is expected to be weak, as \citet{Meeus2001} did not detect any PAH features at 8.6$\,\mu$m in \textit{ISO} data. We therefore expect the flux in the PAH1 band to be dominated by the continuum emission. The emission in both bands is well inside the radius where spiral arms have been found and this suggests that HD 100453 follows the other targets in the sample in which the mid-IR emission is dominated by the central regions. Since the outer disk starts at 17\,AU, the PAH emission seems to come from inside the gap and the NEAR band emission includes the inner wall of the disk which is heated by the star, similar to what is seen in HD 169142.

\subsubsection{HD 36112}
HD 36112 has a large cavity, with an outer disk that has rings, clumps, and spiral arms \cite[e.g.][]{Dong2018,Wagner2019}. However, in our observations, the cavity is unresolved. Since the cavity has a radius of 0.2$\arcsec$ and the upper limit for the 95\% flux radius is 0.07$\arcsec$, this means that most of the emission comes from inside the cavity and not from the inner rim of the outer disk, unlike the NEAR filter emission of the other sources.

\section{Companions}
\label{sec:companions}

\begin{figure*}
    \centering
    \includegraphics[width=\textwidth]{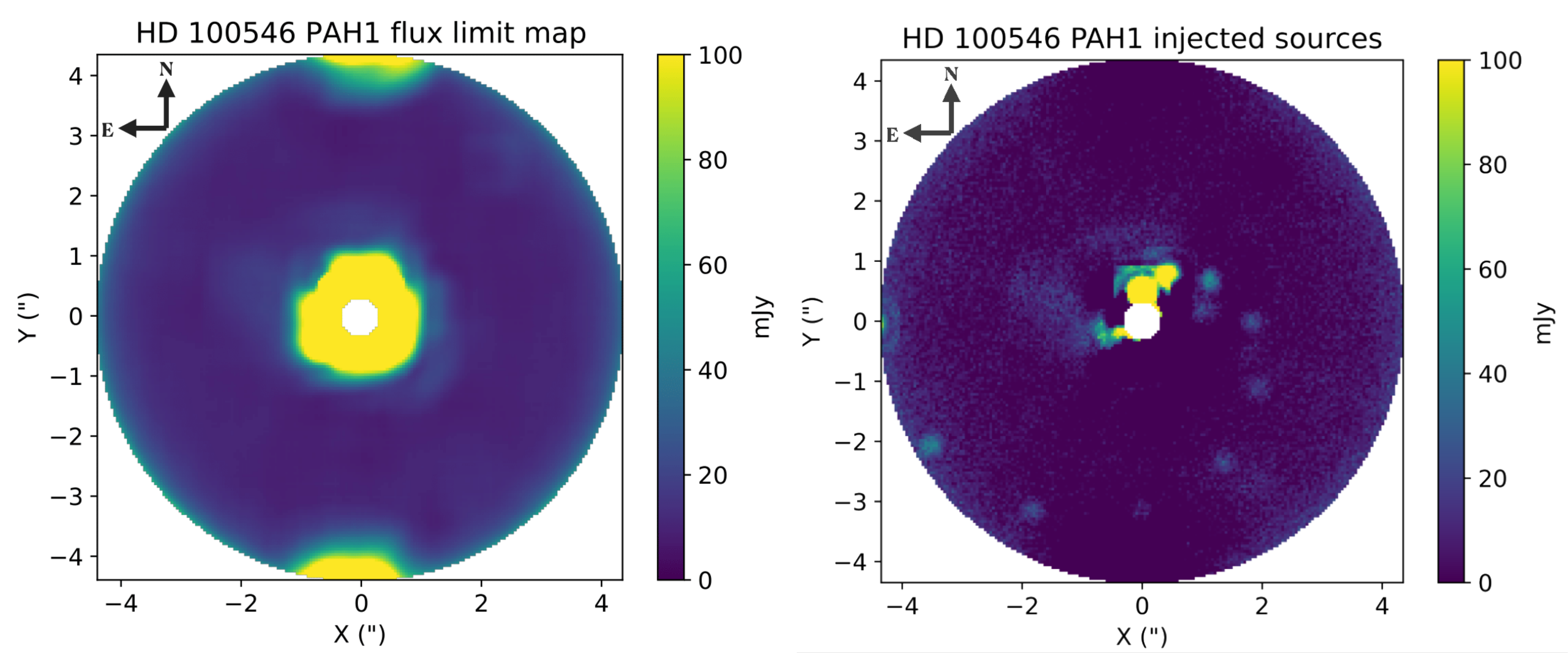}
    \caption{\textit{Left:} Mapped 5$\sigma$ flux limits of the HD 100546 PAH1 data, where the disk image is the most elliptical. The shape of the emitting region does not significantly influence the flux limits, especially beyond 1$\arcsec$ where the data are background limited. \textit{Right:} HD 100546 PAH1 data with sources injected at different separations and position angles at 5 sigma. Most of the sources are clearly visible.}
    \label{fig:fluxmap}
\end{figure*}

\begin{figure*}
    \centering
    \includegraphics[width=\textwidth]{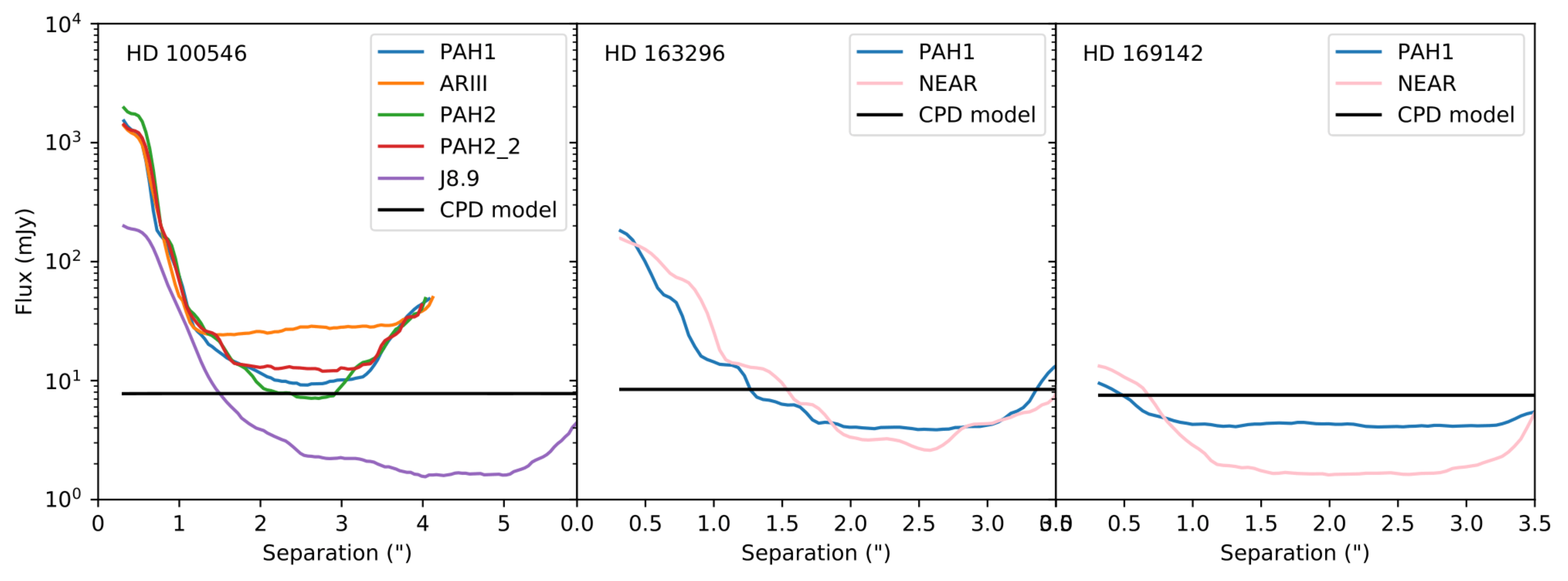}
    \caption{5$\sigma$ flux limits of potential companions to three targets compared to the CPD model described in Table \ref{table:cpd_grid} inserted in the circumstellar disk. \textit{Left:} limits for the different observations of HD 100546 in PAH1 (blue), ARIII (orange), PAH2 (green), PAH2\_2 (red), and J8.9 (purple) filters. The black line indicates the estimated flux as a function of radial separation for our fiducial CPD model as described in column 2 of Table \ref{table:cpd_grid}. Only one line is included as the model values are similar across the different filters. The increase at 7$\arcsec$ in the J8.9 filter and at 4$\arcsec$ in the other filter are the results of chopping and nodding shadows.  \textit{Middle:} Limits for the observations of HD 163296 in the PAH1 (blue) and NEAR (pink) filters, along with the expected flux of the same CPD in the HD 163296 disk. \textit{Right:} The same as the middle figure, but for HD 169142.  }
    \label{fig:limits}
\end{figure*}

\begin{figure}
    \centering
    \includegraphics[scale=.6]{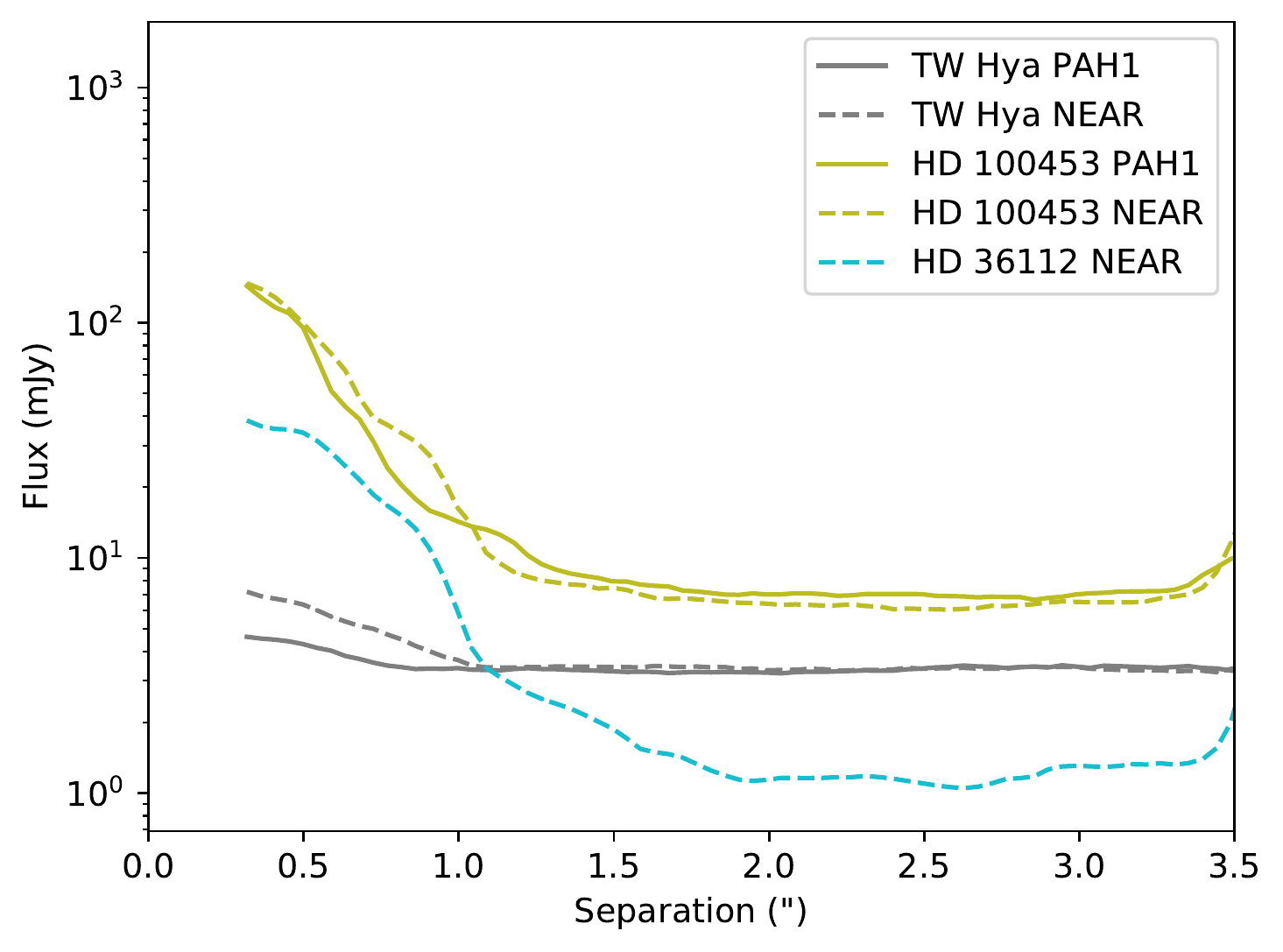}
    \caption{Observational limits on potential companions to TW Hya (grey), HD 100453 (yellow-green), and HD 36112 (turquoise) in the PAH1 (solid lines) and NEAR (dashed lines) filters. The increase at 4$\arcsec$ is the result of shadows from the chopping and nodding in the observations.}
    \label{fig:limits2}
\end{figure}

The proposed companions of the disks in this sample are potential hosts to circumplanetary disks, which thus far have only been tentatively identified in the PDS 70 system \citep{2019A&A...625A.118K, 2019ApJ...877L..33C, Haffert2019, Isella2019}. To search for planetary companions and associated dust concentrations in the disk, the circularised PSF subtraction described in \cite{Petit2019} was applied to the data. This method creates an individual reference PSF from the data for every nod-subtracted image by azimuthally averaging it. The resulting rotationally symmetric PSF was then subtracted from the original data to remove the radially dependent stellar flux. This was decided upon because there is not sufficient rotation in the images to do angular differential imaging and most of the data do not have reference stars available for standard PSF subtraction. Standard PSF subtraction would also not subtract any spatially extended disk emission. Subtracting a circularly symmetric PSF from an elliptical disk image does leave residuals, but the bulk of the disk emission (>80\%) is subtracted. Additionally, the sizes of the emitting regions in our data are small and the flux limits are not influenced by their shapes beyond the very inner pixels, where the disk is visible. This is the case even within roughly 1$\arcsec$, beyond which the background dominates and the shape of the emitting region becomes irrelevant. An example of this can be seen in the left panel of Fig \ref{fig:fluxmap}, where the limits are mapped for HD 100546 in the PAH1 filter, which has the most elliptical image of our entire dataset. While none of the proposed companions are detected in any of the disks, it is possible to set 5$\sigma$ upper limits on the fluxes of any possible companions, based on the residual noise at each possible location. A limit of 5$\sigma$ was chosen, since injected 5$\sigma$ sources were clearly recovered in the reduced data, as can be seen in the right panel of Figure \ref{fig:fluxmap}. The only exception is the source directly to the south of the star, which, although still present, is less clear due to its proximity to one of the shadows induced by the nodding. However, the affected areas around these shadows are small.  

Fig. \ref{fig:limits} and \ref{fig:limits2} show the resulting flux limits, with Fig. \ref{fig:limits} including the flux of a model planet with a circumplanetary disk, which is discussed in the next section. The obtained limits are of the order of a few millijanskys between 1$\arcsec$ and 3.5$\arcsec$ separation up to a few tens of millijanskys at 0.5$\arcsec$. This is more sensitive than any previous mid-IR imaging observations by a factor of 10-100. Beyond 3.5$\arcsec,$ the limits are dominated by the shadows induced by the chopping and nodding procedure in the observations. The differing sensitivities between objects with the same integration times are the result of different observing conditions influencing the data quality of the different targets.

\subsection{Companion models}

The presence of planetary accretion and a CPD or circumplanetary dust envelope can act to significantly increase the mid-IR luminosity of a putative companion \citep[e.g.][]{Zhu2015}. To determine our own mid-IR observational limits for the planet candidates with accompanying CPDs, we explored a grid of CPD models using \textsc{ProDiMo}.  Our model grid consists of a range of possible planet CPD masses, CPD dimensions, dust grain size distributions, and planet luminosities. 

\subsubsection{Properties of the planet and CPD models}

We consider planetary masses of 1 to 10\,$\mathrm{M_{\rm J}}$, with correspondingly sized CPDs defined by the planet's Hill radii. As CPDs could be tidally truncated to $\sim$1/3 of this radius \citep{1998ApJ...508..707Q, Martin2011}, or even photoevaporatively truncated to 0.1-0.16\,$\mathrm{R_{\rm Hill}}$ \citep{Mitchell_2011, Oberg2020}, we set our CPD surface density tapering radius to the point at which the surface density begins to decline exponentially at $\mathrm{R_{\rm Hill}}/3$ and the outer radius at $\mathrm{R_{\rm Hill}}$.   

We considered a range of CPD masses relative to the planet masses $M_{\rm CPD} = 10^{-4}-10^{-2} M_{\rm p}$, and a range of planetary luminosities corresponding to various stages of accretion such that $L_{\rm p} = 10^{-6}-10^{-2}\,\mathrm{L_{\odot}}$ \citep{Mordasini2012}.  \citet{Marley2007} found that a 10\,$\mathrm{M_{\rm J}}$ planet in a 'hot start' evolution scenario can decline monotonically in luminosity from an initial $\sim 4\times 10^{-3}\,\mathrm{L_{\odot}}$ to $\sim 4\times 10^{-4}\,\mathrm{L_{\odot}}$ within 5 Myr.  In the core accretion case, they found a peak luminosity during runaway accretion of $> 10^{-2}\,\mathrm{L_{\odot}}$ which lasts $\sim 3 \times 10^5$ yr, rapidly declining to $\sim 2 \times 10^{-6}\,\mathrm{L_{\odot}}$ by 3 Myr.  Given that the planetary luminosity is expected to peak only briefly at or above $L_{\rm p} \sim 10^{-2}\,\mathrm{L_{\odot}}$, we consider the case of $L_{\rm p} = 10^{-2}\,\mathrm{L_{\odot}}$ to be the most optimistic detection scenario.

Pressure bumps at gap edges are suspected to act as filters for dust grain size, preventing the accretion of grains significantly larger than 10\,$\mu$m onto planets within the gap \citep{Rice2006}.  We thus also considered CPDs where the dust grain size population is limited to maximum sizes of 100 and 10\,$\mathrm{\mu}$m. 

A companion orbiting within an optically thin region of the circumstellar disk can be exposed to significant UV radiation from its host star \citep{Oberg2020}.  Photons of energy 6-13.6\,eV are known as FUV and can efficiently heat disk surfaces. The significant FUV luminosity of the host star can act to heat the surface of the CPD and increase its IR luminosity.  We parameterised the FUV flux with the Draine field $G_0 = 1.6\times10^{-3}$\,erg cm$^{-2}$ s$^{-1}$, which was integrated from 6-13.6\,eV \citep{1968BAN....19..421H}.  We extracted the G\textsubscript{0} field intensity using \textsc{ProDiMo} from the results of the 2D radiative transfer within the DIANA circumstellar disk models and applied this as a UV background field to our own CPD models. Given that dust is the dominant source of opacity in the UV, it should be noted that the gaps in the DIANA disk models (see Fig.\ref{fig:rhod} for the HD100546   dust structure) are free of dust and do not contribute to the UV opacity.

\subsubsection{Companion flux estimates}

We extracted the planet and CPD flux from the SEDs produced by the  \textsc{ProDiMo} continuum radiative transfer and weighed it across the filter response curves. This flux represents the idealised total flux emitted by the unresolved companion, unconvolved with the observational PSF. We find that for high planetary luminosities ($>10^{-4}\,\mathrm{L_{\odot}}$), the mid-IR flux is dominated by the planet itself, whereas the CPD only contributes $3-6\%$ of the combined emission largely independent of CPD properties.  

\begin{figure}
    \centering
    \includegraphics[scale=0.6]{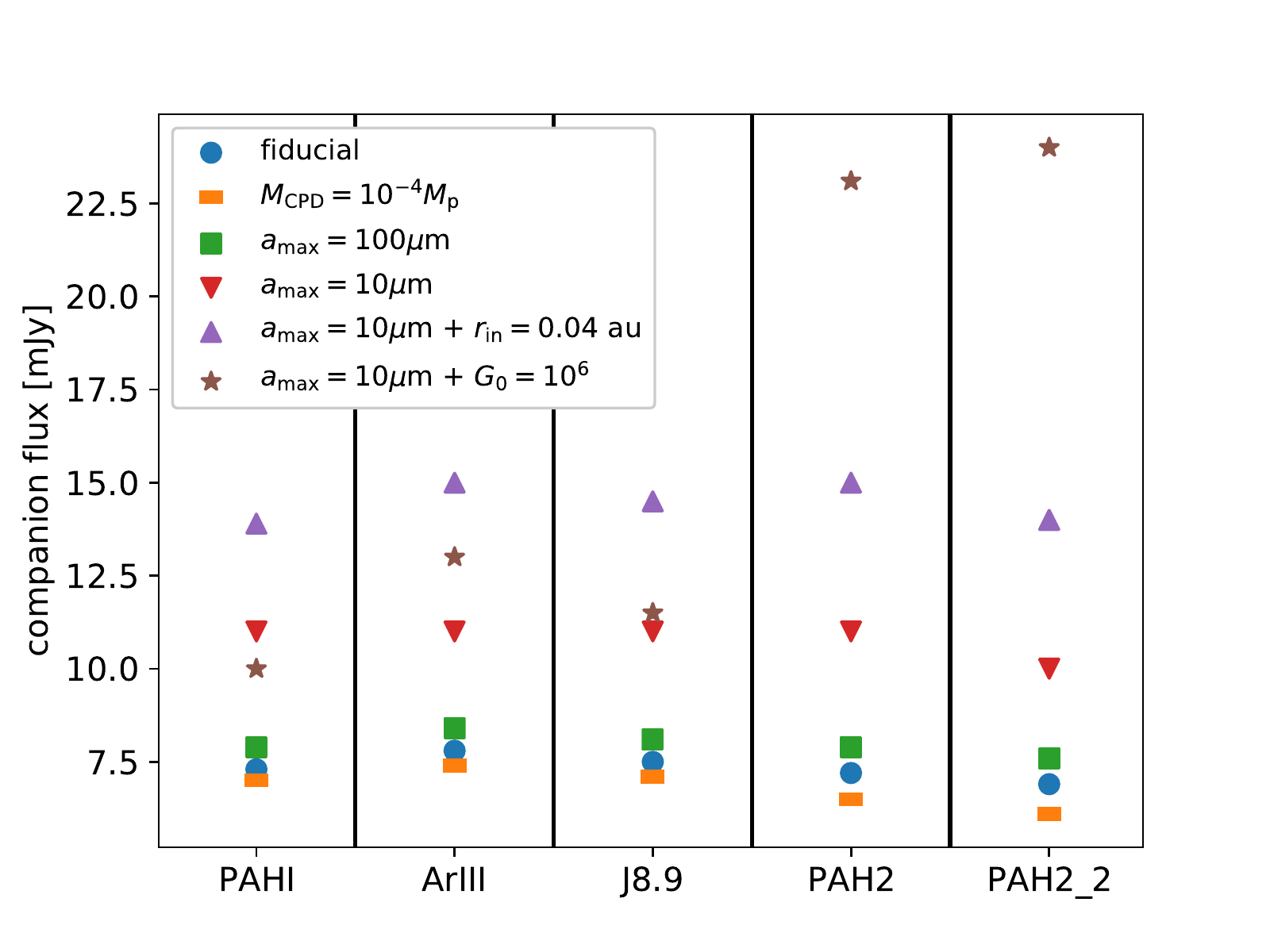}
    \caption{Model companion (planet and CPD) unextincted flux estimates.  The `fiducial' case is described by the planet and CPD parameters found in column 2 of Table \ref{table:cpd_grid} at 55\,AU for the HD 100546 system.  We also consider a variety of maximum dust grain sizes $a_{\rm max}$, CPD mass $M_{\rm CPD}$, CPD inner radius $r_{\rm in}$, and background FUV radiation field strength $G_0$.} 
    \label{fig:cpd_filters}
\end{figure}

For our disk models, the size of the CPD as estimated by its Hill stability and the strength of the background FUV field both vary in predictable ways.  For a given CPD model, our parameter grid exploration thus allowed us to fit for the resulting planet and CPD flux given an arbitrary radial separation from the star. As the vertical dust opacity at arbitrary wavelengths was also calculated as part of our model radiative transfer for various circumstellar disks, we were able to determine the radial dependence of the extinction to the midplane as well. We solved for the dust column density as a function of the viewing inclination for each radial position in the disks, and from this we derived the resulting $9\,\mu$m optical depth. The black line in Fig. \ref{fig:limits} represents the resulting expected flux of the planet and CPD model in the J8.9 filter for a $10\,\mathrm{M_{\rm J}}$ planet with a CPD of mass $10^{-2} \mathrm{M_{\rm p}}$ as described in Table \ref{table:cpd_grid}. The line was derived from a fit performed to the J8.9 flux of our model grid of CPDs in which the background FUV radiation field, the disk size, and extinction to the midplane were simultaneously varied as a function of radial separation, although the predicted flux is relatively flat for planets found outside of the optically thick regions of the circumstellar disks. For low radial separations, the background FUV field heats the CPD surface and results in increased mid-IR emission.  The CPD size grows with increasing distance from the star as the companion's Hill sphere increases correspondingly; however, as the majority of the CPD mid-IR emission originates from the innermost regions of the CPD, this contribution becomes negligible at large separation.  The flux of our CPD models in the other filters is similar, varying for non-pathological model cases by at most $\sim$10\%, and they are thus roughly comparable, as illustrated in Fig. \ref{fig:cpd_filters}. 

\subsubsection{Results for HD 100546}

While previous estimates of the age of HD 100546 indicate an older ($\sim 10$\,Myr) system \citep{Ancker1997}, \citet{Fairlamb2015} derived an age of $7.02 \pm 1.49$\,Myr and an accretion rate of $\dot M \approx 10^{-7}\,\mathrm{M_{\odot}}$ yr$^{-1}$.  The mass of the HD 100546 inner disk was fit to be $8.72\times10^{-8}\,\mathrm{M_{\odot}}$ \citep{Woitke2019}, thus requiring continuous replenishment from the outer zone across the gap.  The plausibility of an actively fed circumplanetary accretion disk is thus supported by the ongoing presence of radially evolving dust within the circumstellar disk \citep{Marley2007,Mordasini2012}.  

We considered companions placed in the midplane at multiple radial separations from the star to study the influence of the background radiation field and circumstellar dust extinction on the predicted flux.  We considered the properties of the  planet candidate HD 100546b described by \citet{Quanz2015}, which was found at a radial separation of 53$\pm2$\,AU.  When the planet was treated as a single-temperature blackbody, \citet{Quanz2015} found the best fit solution to be an emitting region of $R = 6.9^{+2.7}_{-2.9} R_{\rm J}$ with $T = 932^{+193}_{-202}$ for a luminosity $L = 2.3^{+0.6}_{-0.4} \times 10^{-4}L_{\odot}$.  As the addition of a CPD may produce an emission signature diverging significantly from a single-temperature blackbody, we loosened the constraints on the temperature and emitting area. For a 2.5\,$\mathrm{M_{\odot}}$ star, a planet of 1, 5, or 10\,$\mathrm{M_{\rm J}}$ at 55\,AU has a Hill radius  of 2.77, 4.73, or 5.96\,AU, respectively.   We considered three cases in detail: a planet immediately interior to the outer gap wall at 18\,AU,  a planet embedded within the outer gas and dust disk at 55\,AU, and a wide-separation planet in the optically thin region of the PPD at 100\,AU, with correspondingly sized CPD outer radii, maximum dust grain sizes, FUV backgrounds, and optical depths to the midplane (see Table \ref{table:cpd_grid}).  While the CPD size, as set by the Hill radius,  only varies by a factor of 100 across the disk surface from 5-500\,AU, the background UV radiation field varies more dramatically by a factor $> 10^6$.

At the radial location of the 55\,AU planet candidate, we extracted  an FUV flux
of $G_{0} = 10^{3.65}$ in the midplane from the results of our circumstellar disk model radiative transfer. At 5\,AU in the shadow of the inner disk, we find $G_{0} = 10^{5.4}$ and at 18\,AU $G_{0} = 10^{6.5}$.   The maximum $G_0$ within the gap is found to be $~3\times10^6$.  The gas component of a CPD experiencing such irradiation acquires an optically thin heated envelope with a temperature of around 5000\,K at $z/r \sim 0.4$.  The $\sim$70\,K optically thick surface below this envelope gives rise to significant re-radiated emission peaking at 30-50\,$\mathrm{\mu m}$.  The short-wavelength tail of this component contributes non-negligibly to the J8.9 flux across the entire CPD surface for $G_0 > 10^5$.   

From the HD 100546 disk model dust density distribution and dust opacities, we determined the optical depth to the midplane along the line-of-sight to the observer across the J8.9 band to determine extinction at arbitrary radii.  While emission arising from planets inside the gap would be largely unextincted, immediately outside of the gap we find a maximum optical depth $\tau_{\rm J8.9}$ of 5.6.  The disk becomes optically thin at 8.7\,$\mathrm{\mu}$m only outside of 82\,AU. We find that at the large separations where our sensitivity is maximal at $a > 160$\,AU, $\tau_{9 \mu m}$ is at most 0.18 and $\tau \propto a^{-2.4}$. 

The model planet with a mass of 10\,$\mathrm{M_{Jup}}$ and a luminosity of $10^{-2}\,\mathrm{L_{\odot}}$ would have been detected in the J8.9 data beyond this radius and in PAH2 between 2$\arcsec$ and 3$\arcsec$. Hence, our new mid-IR imaging data prove that no such massive, luminous planets exist in the HD 100546 system at radii larger than 160\,AU from the central star.  A companion with a luminosity of $10^{-3}\,\mathrm{L_{\odot}}$ would be marginally detectable at angular separations of 4-5$\arcsec$ only.

\subsubsection{Results for other systems}

We used a single best-case representative planet and CPD to derive detection limits for the other observed systems as a function of separation. The model CPD mid-IR flux levels are constant at  large radii, because at large separations the UV radiation emitted by the star no longer significantly contributes to the heating and re-radiation of the CPD. The fact that the CPD is free to physically increase in size as the planet's Hill radius increases also no longer acts to increase the flux, as for the optically thick CPDs we consider, the planet acts only to heat the innermost regions of the CPD, from which the majority of the 9\,$\mathrm{\mu}$m emission originates.  

For HD 163296, we excluded a 10\,$\mathrm{M_{Jup}}$, $10^{-2}\,\mathrm{L_{\odot}}$ companion between 1.5$\arcsec$ and 3.5$\arcsec$, as it would have been observed in both filters. For HD 169142, TW Hydra, and HD 36112, we excluded it beyond 1$\arcsec$. HD 100453 is the only system in which it would remain undetected.

\subsubsection{Reconciling prior observational constraints}

In previous work, the planet candidate HD 100546 b at 55\,AU separation is the only companion that has had its putative CPD constrained in mass to 1.44\,$\mathrm{M_{\oplus}}$ (or $2.7 \times 10^{-3}\,\mathrm{M}_{\rm p}$ for a planet mass 1.65\,$\mathrm{M}_{\rm Jup}$) in the optically thin case, and a size of 0.44\,AU in radius for the optically thick case, although this rests on assumptions regarding the grain size population of the CPD and the ratio between planetary and CPD mass \citep{2019ApJ...871...48P}. ALMA observations of HD 100546 at 870\,$\mathrm{\mu}$m set a 3$\sigma$ limit of 198\,$\mathrm{\mu}$Jy for any planet candidate \citep{2019ApJ...871...48P} with which we can further constrain any CPD's longwave emission. 

We find that for our fiducial CPD surrounding a 10\,$\mathrm{M}_{\rm Jup}$ planet of $10^{-2}\,\mathrm{L_{\odot}}$, we overpredicted the upper limit set by ALMA observations at 870\,$\mathrm{\mu}$m by a factor of 13.  When the fiducial CPD is modified with a maximum grain size of $10\,\mathrm{\mu}$m, this overprediction is reduced by a factor of $\sim$2.  Our planet and CPD models can be brought into agreement with the ALMA flux limits by reducing the mass of the CPD relative to the planet or by reducing the dust-to-gas ratio.  We find that while the 9\,$\mathrm{\mu}$m flux of the CPDs is largely insensitive to their mass, the continuum flux in ALMA band 10 is primarily dependent on our CPD mass, radius, and dust-to-gas ratio owing to the emission region corresponding to cooler dust at larger separation from the planet \citep{Rab_2019}. For a fixed radius, dust-to-gas ratio, maximum and minimum dust grain size, and grain size power law, the 870\,$\mathrm{\mu}$m flux is proportional to the CPD mass as $F_{870 \mu m} \propto M_{\rm CPD}^{0.81}$  for the range $M_{\rm CPD} = 10^{-6}-10^{-2}\,\mathrm{M}_{\rm p}$.  We find that the maximum CPD mass allowed by the constraint is $3.2 \times 10^{-7}\,\mathrm{M_{\odot}}$.  A smaller, optically thick CPD of a higher mass still satisfies the constraint.  We find that a modification to our fiducial CPD of a mass $>9.5\times10^{-6}\,\mathrm{M_{\odot}}$ with a tapering radius of 0.2\,AU and an outer radius of 0.6\,AU has a 870\,$\mathrm{\mu}$m flux of 190\,$\mathrm{\mu}$Jy and would thus satisfy the constraint set with ALMA.  This places no additional constraints on our 9\,$\mathrm{\mu}$Jy flux prediction, as the mid-IR flux is instead primarily dependant on the planet's luminosity and the CPD's inner radius.

\begin{table*}
    \centering
    \caption{HD100546 candidate planets and CPD model parameters for our optimistic detection scenario (parameters listed above the first horizontal divider) for a variety of radial separations (parameters below the first horizontal divider) and associated J8.9 band predicted fluxes. Dust composition is identical to that listed in Table \ref{table:disk_params}.}
   \begin{tabular}{ll|l|l|l}
    \hline \hline
        Parameter                       & Symbol      & 18 AU &55 AU & 100 AU       \\ \hline
        Planet Mass [$M_{\rm J}$ ]   & $M_{\rm p}$        & 10           & 10      & 10         \\
        Planet Luminosity [$L_*$]    & $L_{\rm p}$        & $10^{-2}$    &$10^{-2}$& $10^{-2}$  \\
        CPD Mass    [$M_{\rm p}$]       & $M_{\rm CPD}$      & $10^{-2}$    &$10^{-2}$& $10^{-2}$  \\
        
        CPD Inner Radius   [AU]         & $R_{\rm CPD,in}$   & 0.01         & 0.01    & 0.01       \\

        Minimum Dust Size [$\mu m$]     & $a_{\rm min}$      & 0.05         & 0.05    & 0.05       \\
        
        Col. Density Power Index      & $\epsilon$      & 1         & 1    & 1       \\

        Dust to Gas Ratio   & $d/g$      & 0.01         & 0.01       & 0.01          \\
        
        Reference Scale Height   & $H_{\rm 0.1 au}$      & 0.01         & 0.01       & 0.01          \\
         
        \hline
        Planet Semi-major Axis [AU]   & $a_{\rm p}$        & 18           & 55      & 100        \\
        
        CPD Tapering Radius   [AU]               & $R_{\rm tap, CPD}$      &  0.40        & 1.99    &  3.61      \\

        CPD Outer Radius   [AU]               & $R_{\rm out, CPD}$      &  1.19         & 5.96    &  10.84      \\

        Maximum Dust Size [$\mu m$]     & $a_{\rm max}$      & 10         & 3000      & 3000        \\
 
        FUV background          & $G_0$              &    $10^{6.7}$          &  $10^{4.1}$       & 3500  \\
        
        Optical depth at 8.7 $\mu$m         & $\tau$              &      $\sim 0$    &  1.27       & 0.43  \\

        \hline

        \begin{tabular}[c]{@{}c@{}} Predicted 8.7\,$\mu m$ flux (extincted) [mJy]\end{tabular}&$F_{\rm P,1}$    &  11.5     &  2.1  &  4.88 \\
        
        \begin{tabular}[c]{@{}c@{}} Predicted 8.7\,$\mu m$ flux (unextincted) [mJy]\end{tabular}&$F_{\rm P,0}$    &  11.5     &  7.6  &  7.5 \\
        \hline
        
    \vspace{1ex}
    \end{tabular}
    
    \label{table:cpd_grid}
\end{table*}

\noindent

\section{Discussion and conclusions}
\label{sec:conclusions}

We analysed images of HD 100546 in five different mid-IR filters and a further five young stellar objects in the PAH1 and NEAR infrared filters with the VISIR instrument and its upgrade NEAR. The resolved disks had their FWHMs and inclinations determined. HD 100546 has a FWHM of 28-61\,AU across five different filters, a projected inclination of 44$\pm4^\circ$, and a projected position angle of 130$^\circ$. HD 169142 has FWHMs of 29\,AU and 41\,AU in the PAH1 and NEAR filter bands, respectively, and a projected inclination of 13$\pm2^\circ$. HD 100453 has a FWHM of 9\,AU in the PAH1 band and 21\,AU in the NEAR band and an inclination of 35$\pm2^\circ$.
The observed values are consistent with the DIANA circumstellar disk models and previous observations of the sources. We set upper limits of 6\,AU and 7\,AU on the size of the emission region of HD 163296 in the PAH1 and NEAR filter bands, respectively, thus improving previous limits by a factor of three. We set upper limits of 3\,AU and 7\,AU on TW Hydra in the same filters, which is consistent with previous observations. Finally, we set an upper limit of 13\,AU on the size of the NEAR filter emission of HD 36112, which is an improvement over previous values of a factor of 10. The fact that we did not resolve these targets is also consistent with the DIANA \textsc{ProDiMo} models \citep{Woitke2019}. Because of the method by which the variety of observational data were weighted during the original fitting procedure performed by \citet{Woitke2019}, and because of the non-complete set of disk model parameters for which the fits were performed, localised improvements to the SED were still possible. After a minimal adjustment of the HD 100546 disk model gap geometry, an examination of the disk radial profile showed that our \textsc{ProDiMo} model was a good match for the data and that it reproduces the radial profile of the disk to within 1$\sigma$ without the need to include a companion object.  In all cases, the mid-IR emission originates from the central area of the disk from the most highly irradiated areas: unresolved inner disks and/or the inner rims of the outer disks. 

Given our new flux estimate for the HD 100546 system, we have improved upon the global SED fit from 2-18\,$\mathrm{\mu}$m by simultaneously increasing the gap outer edge from 19.3\,AU to 22.3\,AU, increasing the abundance of PAHs in the outer disk relative to the ISM from 2.8$\times10^{-3}$ to 3.4$\times10^{-3}$, and replacing the representative PAH circumcoronene with coronene. The details of the PAH properties fitting can be found in Appendix \ref{appendix:PAH_fitting}. Given that the spectral properties of alternative dust compositions have not been thoroughly explored nor the marginal improvement of the detailed PAH fit, we tend to favour the simple modification of only the disk gap geometry. The $\chi^2$ statistic between the model SED and the \textit{ISO-SWS} spectrum for 2-18\,$\mathrm{\mu}$m reduces from 588 to 278 when the inner radius is increased to 22.3\,AU.  It should be noted that increasing the model gap outer radius would act to increase the tension with the location of the dust continuum gap edge observed with ALMA at 16-21\,AU \citep{Perez2020}, although as ALMA traces millimetre-sized grains, this may not be inconsistent. Additionally,  the model gap outer radius is the one parameter that we adjusted which was previously fit by means of a genetic algorithm \citep{Woitke_2016,CDAG2,Woitke2019, CDAG4}. 

We produced planet and CPD flux estimates using the thermochemical disk modelling code \textsc{ProDiMo} for the VISIR filters with a variety of CPD parameters, finding that in the absence of extreme external FUV radiation fields, the maximum unextincted flux in the J8.9 band is expected to be $\sim$15\,mJy for a CPD with an inner radius of 0.04\,AU and a maximum dust grain size of 10\,$\mathrm{\mu}$m.  We find that this flux is largely dependent on the planet properties and not on those of the circumplanetary disk.  The CPD is found to contribute $3-6\%,$ at most, of the companion flux at 9\,$\mathrm{\mu}$m. The CPD contribution at 9\,$\mathrm{\mu}$m is greatest when the maximum grain size is reduced to 10\,$\mathrm{\mu}$m and the CPD is irradiated by a significant FUV field of $G_0 \geq 10^6$. 

Such conditions are found within the gap of the HD 100546 disk, where we determined that the $G_0$ field strengths up to $3\times10^6$, despite the presence of the inner disk. A planet and CPD within the gap at 18\,AU, while more gravitationally truncated than our test cases at 55 and 100\,AU, is unobscured by dust and we expect $F_{\rm J8.9} = 11.5$\,mJy.  We note that while the 9\,$\mathrm{\mu}$m emission of the CPD is largely unaffected for $G_0 \leq 10^6$, it rises precipitously above this, and for a $G_0 = 10^7$ we find $F_{\rm J8.9}  = 0.6$\,Jy. While a CPD within the gap would be found at angular separations of less than 0.2$\arcsec$ and thus be unresolved in our observation, the contribution to the flux of the star and circumstellar disk (31 $\pm$ 3\,Jy) would thus be non-negligible.   It should be noted however that a significantly FUV irradiated CPD can become photoeveporatively truncated such that the effective emission region is greatly reduced \citep{Oberg2020}.  

For our $a = 55$\,AU HD 100546 companion test case, we find $F_{\rm J8.9}  = 2.1$\,mJy owing to significant dust extinction. In the event that the planet is able to clear obscuring dust from its immediate neighbourhood in a localised cavity, the observed flux may increase to 7.5\,mJy. Even in this 'best case' scenario of high planetary luminosity, it can be seen in Fig. \ref{fig:limits} that the flux limiting sensitivity at 55\,AU is 200\,mJy.  For our $a = 100$\,AU companion case, we find  $F_{\rm J8.9}  = 4.9$\,mJy, 7.9\,mJy unobscured, and the accompanying limiting sensitivity is 30\,mJy.  Only outside of 180\,AU would such a planet and CPD be detectable.  Outside of 180\,AU, we find a limit on planetary luminosity of 0.0028\,$\mathrm{L_{\odot}}$, above which we would have detected any companion. 

In the HD 100546 system, we rule out the presence of planetary mass companions with $L >  0.0028 \mathrm{L_{\odot}}$ for $a > 180$\,AU. We find that the contribution of a planet and CPD would still be of the order of the uncertainties inherent in the model, as relatively minor modifications to the HD 100546 gap dimensions (an increase of 2-3\,AU in the outer radius) produce changes in expected continuum flux of 7-10\,Jy at $9\,\mathrm{\mu}$m. We place no stringent constraints on the planetary mass, CPD radius, or CPD grain size distribution. In the HD 169142, TW Hydra, and HD 100453 systems, we can exclude companions with $L > 10^{-2}\,\mathrm{L_\odot}$ beyond 1$\arcsec$.

We consider whether the lack of detection of wide-separation ($a$>50\,AU) planetary mass companions (PMCs) of mass $<\,20\,M_{\rm J}$  in the five studied systems is remarkable. While the presence of a dusty CPD may act to enhance the observability of a companion, it has been found that rapid dust evolution in CPDs of isolated wide-separation PMCs could act to suppress the dust-to-gas ratio of CPDs on short timescales ($d/g\leq10^{-4}$ after 1 Myr), rendering a continuum detection more challenging \citep{pinilla2013, zhu2018, Rab_2019}. Sub-stellar companions have been detected in wide orbits around young stars \citep{Neuhauser2005,Ireland2011,bryan2016,naud2017,bohn2019}.  It has been suggested that such objects may form in situ by the fragmentation of massive,  self-gravitating disks \citep{Boss1997,Boss2011,Vorobyov2013} by the direct collapse of molecular cloud material \citep{Boss2001}, or by core- or pebble accretion \citep{Lambrechts2014} and subsequent outwards scattering by an interaction with other massive planets \citep{Pollack1996,Carrera2019}.  In the latter case, a detection of a wide-separation PMC may thus directly imply the presence of additional massive planets in the inner system.

\citet{Bowler2016} suggests that around single, young (5-300\,Myr) stars, 5--13\,$M_{\rm J}$ companions at separations of 30-300\,AU occur $0.6^{+0.7}_{-0.5}\%$ of the time.  With VLT/NaCo, \citet{Vigan2017} found that 0.5-75\,$M_{\rm J}$ companions at separations of 20-300\,AU are found around $0.75-5.7\%$ of stars, and with the Gemini Planet Imager Exoplanet Survey, \citet{Nielsen2019} found that  5-13\,$M_{\rm J}$ companions with separations of 10-100\,AU occur around $9^{+5}_{-4}\%$ of stars. Our sensitivity at the limiting angular resolution restricted our search to relatively wide separation companions ($a$ > 160\,AU). Given the PMC occurrence rate of \citet{Bowler2016}, we expect an absolute upper bound of $\sim3.4^{+3.9}_{-3.3} \%$ probability of a single detection in our sample, assuming a perfect detection efficiency from 30-300\,AU.  In this context, it is difficult to make new conclusions regarding the prevalence of wide-separation PMCs in our observed systems given the relatively low a priori likelihood of detection and the relatively large companion luminosity ($10^{-3}-10^{-2}\,L_{\odot}$) necessitated. We were able to set an upper limit to the occurrence rate for wide-separation PMCs with a luminosity $\geq 10^{-2}$ of $\leq 6.2\%$ at $68\%$ confidence.

Future observations with METIS \citep{Brandl2018} on the ELT are expected to achieve ten times better sensitivities than NEAR and 40 times better sensitivities than VISIR at the same wavelengths, as well as improving the spatial resolution by a factor of 5, allowing for one to  image more close in companions. MIRI\footnote{\url{https://www.stsci.edu/jwst/science-planning/proposal-planning-toolbox/sensitivity-and-saturation-limits}} on \textit{JWST} is expected to achieve 250 times better sensitivities than NEAR and 1000 times better sensitivities than VISIR. Both will be able detect the known companions to all six targets.

\begin{acknowledgements}
    This project was made possible through contributions from the Breakthrough Foundation and Breakthrough Watch program, as well as through contributions from the European Southern Observatory.  
\end{acknowledgements}

\bibliographystyle{aa}
\bibliography{bibliography}

\begin{appendix}

\section{Standard disk models and SED fitting methodology} \label{appendix:SED_fitting}

To perform the SED fits, a comprehensive set of publicly available observational data, consisting of photometric fluxes, interferometric data, low and high resolution spectra, emission line fluxes, line velocity profiles, and maps were used from which the physical and chemical parameters of the disk could be derived (references for which can be found in \citet{CDAG4}). The fits were performed by iteration of parameter sampling in MCFOST radiative transfer models by means of a genetic algorithm.   HD 100546 was fit with 120 data points, two disk zones, PAHs, and 16 free parameters total after 632 generations and 7584 models. Further details of the standard disk models, SED fitting procedures, and the limitations of SED fitting can be found in \citet{Woitke_2016}, \citet{CDAG2}, \citet{Woitke2019}, and \citet{CDAG4}.

\subsection{Limitations}

The DIANA SED fitting procedure was performed with dust opacities corresponding to a mixture of amorphous pyroxene silicates and amorphous carbon \citep[see Table \ref{table:disk_params};][]{Dorschner_1995,Zubko_1996}. Due to the use of standard dust opacities and a fixed PAH morphology, only the power-law of the dust size distribution and volume fraction of amorphous carbon was varied for the fit, so detailed matching of the spectral features is not expected.  The 8.6\,$\mu$m PAH complex feature, associated with in-plane C-H bending modes, is not fit in detail relative to the \textit{ISO-SWS} spectrum. The presence of an unidentified broad feature at 7.9-8 $\mu$m is not explained by the model, but it has been suggested by \citet{Joblin2009} to originate from a PAH population known as PAH$^x$ consisting of compact but large ionised PAHs with $\sim$100 or more carbon atoms not included in our radiative transfer modelling. 

We opted not to explore the parameter space of possible dust compositions to perform a detailed opacity fitting across the mid-IR given that properties such as the amorphous carbon volume fraction can have a large impact on the SED at all wavelengths, such as by changing the millimetre and centimetre slopes \citep{Woitke_2016}. While the mid-IR traces the disk surface, any features may not be indicative of the disk global dust properties and could represent surface effects, for example, PAHs confined to the surface which are generated locally.  In this case, altering global dust properties may not be the correct approach.

We did not re-perform the global SED fitting procedure to account for the increased GAIA EDR3 distance for HD 100546, but we did consider the implications of an increased stellar luminosity to match the observed luminosity and new distance.  To test the sensitivity of the SED to this adjustment, we considered a modest increase in our stellar effective luminosity to 34.74 L$_{\odot}$. If we were then to scale the physical dimensions of the disk and its gap accordingly, the resulting SED would exhibit a net decrease in mid-IR emission;  across the J8.9 band, we find a deficit in emission over the fiducial model of $2.9\%$.  As this falls within our own observational uncertainty, we do not consider the implications of the new distance estimate further.

\begin{figure}
    \centering
    \includegraphics[scale=0.5]{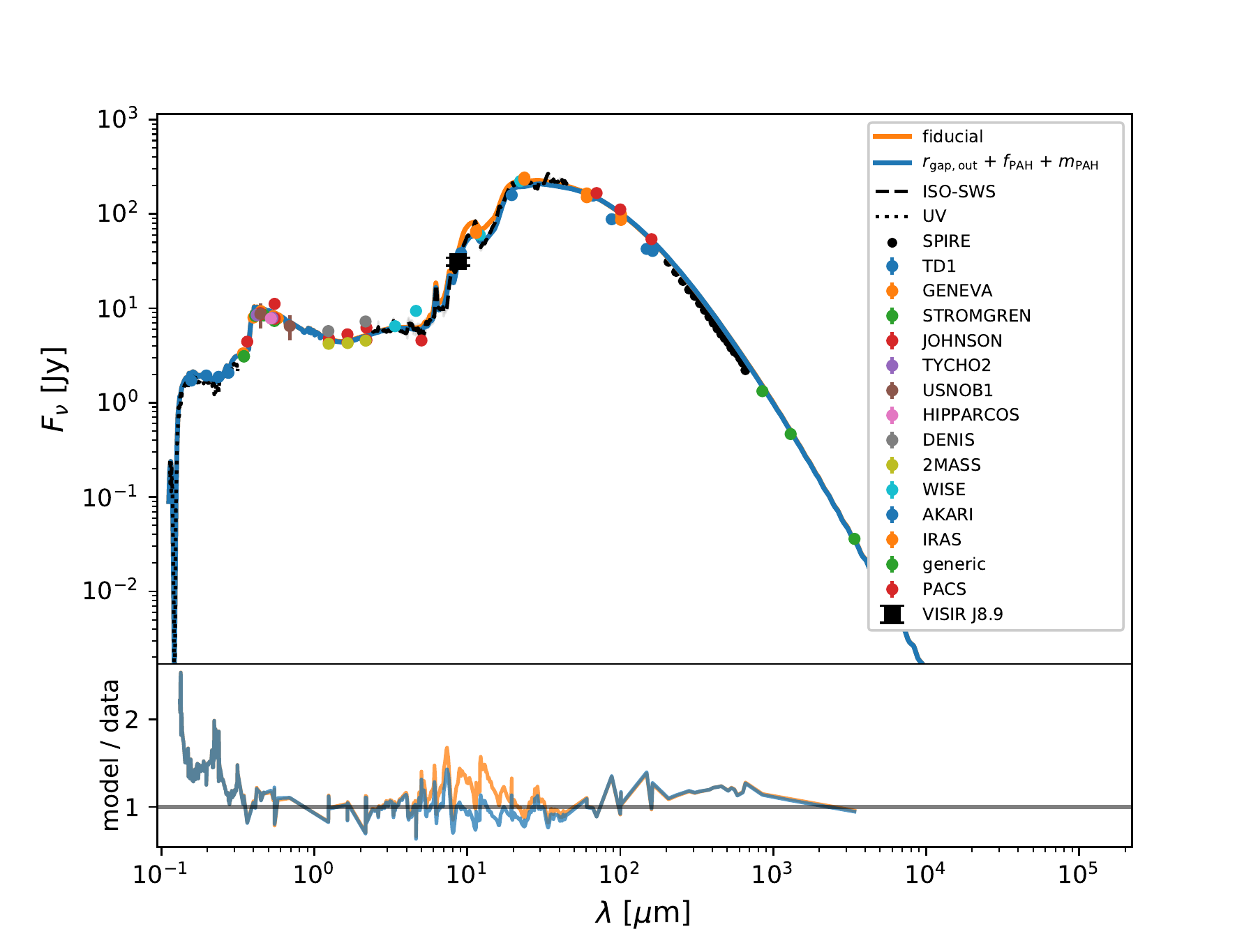}
    \caption{Global SED of the HD100546 disk models and comparison to the observational data folded into the fit. The fiducial model SED is the orange curve and our adjusted disk gap geometry model is the blue curve.  The relative residual as defined by dividing the model by the data is shown at the bottom.}
    \label{fig:global_sed}
\end{figure}

\section{HD 100546 disk model PAH properties exploration} \label{appendix:PAH_fitting}

Several PAH features contribute to the disk opacity near 9\,$\mathrm{\mu}$m.  The broadband filter used in these observations covers an area around 8.6\,$\mathrm{\mu}$m where PAH C-H in-plane bending modes can contribute to the continuum emission.  \textsc{ProDiMo} uses synthetic PAH opacities for neutral and charged PAHs as calculated according to \citet{Li2001}. Exploring the properties of PAHs in the model offers the possibility of modifying the disk flux across the J8.9 filter without globally modifying the disk dust properties and breaking the  quality of the global SED fit.  

The contribution of PAHs was estimated by \citet{vanBoekel2004} to be around 22\% of the total flux near 9\,$\mathrm{\mu}$m. They found the PAH emission to be more extended than the continuum along the spatial dimension of their longslit spectra, with a FWHM of $\approx$150\,AU. Using the low resolution spectroscopic mode of VISIR, \citet{Verhoeff2009} found a statistically significant increase in the spatial extent of the disk emission at 8.6\,$\mathrm{\mu}$m over the resolved continuum emission at a 27 $\sigma$ level.  While they found the ratio between the continuum subtracted peak flux at the 8.6\,$\mathrm{\mu}$m PAH feature over the peak flux was only 2.4$\%$, the deconvolved FWHM size of the continuum subtracted feature was 1.64$^{+0.37}_{-0.38}\,\arcsec$. At a distance of 108\,pc, this corresponds to a disk radius of 178$^{+40}_{-41}$\,AU.  Furthermore, the variability of the 8.6\,$\mathrm{\mu}$m features between \textit{ISO} and TIMMI2 spectra and their respective slit sizes implies that the PAH emitting region is at least 100\,AU in size \citep{Verhoeff2009}. Additionally, \citet{Panic2014} found the 8.6\,$\mathrm{\mu}$m PAH emission to be emitted primarily from angular scales corresponding to $\sim$100\,AU from the star.  

While the HD 100546 disk model PAH abundance and charge fraction was fit for, these parameters were not varied between the inner and outer disk zones.   We thus considered modifications to the PAH population in the outer disk, outside of $r = 22$\,AU, specifically. The DIANA models use a single representative PAH, circumcoronene (C$_{54}$H$_{18}$), and a constant mixture of charged and neutral opacities throughout the disk \citep{Woitke_2016}. For HD 100546, the abundance of PAH relative to the ISM $f_{\rm PAH}$ (defined such that in the ISM $f_{\rm PAH}$ = 1) is 0.0028.   The mean PAH charged fraction is 0.9. We considered both differing PAH types and abundances in the inner and outer disk zones to refine our fit. 

We have explored a grid of a PAH abundance and morphologies in an attempt to minimise the residuals with our mid-IR observational data. Simultaneously allowing for the outer wall of the gap, the abundance of the PAHs, and the type of the PAHs to vary has allowed us to improve upon the standard SED fit without reducing the quality of the fit globally (see Fig.\ref{fig:global_sed}). The result of this multi-parameter exploration can be seen in the green line in Fig. \ref{fig:models}.  We find that a smaller PAH, coronene (C$_{24}$H$_{12}$), and a $22\%$ increase in $f_{\rm PAH}$ outside of the gap wall produce the best agreement with an observation across the J8.9 filter.

\end{appendix}
%
%

%
%
\end{document}